\documentclass[a4paper,UKenglish,cleveref, autoref, thm-restate]{lipics-v2021}

\title{Strategic Liquidity Provision in Uniswap v3}

\author{Zhou Fan}{Harvard University, USA}{zfan@g.harvard.edu}{}{}

\author{Francisco Marmolejo-Cossio}{Harvard University, USA \and IOG, USA}{fjmarmol@seas.harvard.edu}{}{}

\author{Daniel Moroz}{Harvard University, USA}{dmoroz@g.harvard.edu}{}{}

\author{Michael Neuder}{Harvard University, USA}{michael.neuder@gmail.com}{}{}

\author{Rithvik Rao}{Harvard University, USA}{rithvik321@gmail.com}{}{}

\author{David C. Parkes}{Harvard University, USA}{parkes@eecs.harvard.edu}{}{}

\authorrunning{Z. Fan et al.}

\Copyright{Zhou Fan and Francisco Marmolejo-Cossio and Daniel Moroz and Michael Neuder and Rithvik Rao and David Parkes}



\hideLIPIcs

\acknowledgements{This work is supported in part by two generous gifts to the Center
for Research on Computation and Society at Harvard University,
both made to support research on applied cryptography and society.}

\nolinenumbers 

\usepackage{tikz}
\usetikzlibrary{arrows.meta}
\usepackage{pgfplots}
\pgfplotsset{compat=1.17}
\usepackage{graphicx}
\usetikzlibrary{shapes.multipart}
\usetikzlibrary{positioning}
\usetikzlibrary{chains}

\usepackage{neuralnetwork}
\usepackage{listofitems} 
\usepackage[outline]{contour} 
\contourlength{1.4pt}

\tikzset{>=latex} 
\usepackage{xcolor}
\colorlet{myred}{red!80!black}
\colorlet{myblue}{blue!80!black}
\colorlet{mygreen}{green!60!black}
\colorlet{myorange}{orange!70!red!60!black}
\colorlet{mydarkred}{red!30!black}
\colorlet{mydarkblue}{blue!40!black}
\colorlet{mydarkgreen}{green!30!black}
\tikzstyle{node}=[thick,circle,draw=myblue,minimum size=22,inner sep=0.5,outer sep=0.6]
\tikzstyle{node in}=[node,green!20!black,draw=mygreen!30!black,fill=mygreen!25]
\tikzstyle{node hidden}=[node,blue!20!black,draw=myblue!30!black,fill=myblue!20]
\tikzstyle{node convol}=[node,orange!20!black,draw=myorange!30!black,fill=myorange!20]
\tikzstyle{node out}=[node,red!20!black,draw=myred!30!black,fill=myred!20]
\tikzstyle{connect}=[thick,mydarkblue] 
\tikzstyle{connect arrow}=[-{Latex[length=4,width=3.5]},thick,mydarkblue,shorten <=0.5,shorten >=1]
\tikzset{ 
  node 1/.style={node in},
  node 2/.style={node hidden},
  node 3/.style={node out},
}
\def\nstyle{int(\lay<\Nnodlen?min(2,\lay):3)} 

\tikzstyle{mynode}=[thick,draw=blue,fill=blue!20,circle,minimum size=22]

\tikzset{%
  every neuron/.style={
    circle,
    draw,
    minimum size=1cm
  },
  neuron missing/.style={
    draw=none, 
    scale=4,
    text height=0.333cm,
    execute at begin node=\color{black}$\vdots$
  },
}

\usepackage{tabularx}
\usepackage{booktabs}

\newcommand{\matr}[1]{\mathbf{#1}}
\newcommand{\vect}[1]{\boldsymbol{\mathbf{#1}}}

\newcommand\cA{\mathcal{A}}
\newcommand\cB{\mathcal{B}}

\newcommand\cP{\mathcal{P}}

\newcommand\cN{\mathcal{N}}
\newcommand\cS{\mathcal{S}}
\newcommand\cV{\mathcal{V}}

\definecolor{english}{rgb}{0.0, 0.5, 0.0}

\EventEditors{}
\EventNoEds{0}
\EventLongTitle{}
\EventShortTitle{}
\EventAcronym{}
\EventYear{}
\EventDate{}
\EventLocation{}
\EventLogo{}
\SeriesVolume{}
\ArticleNo{25}

\begin{document}

\maketitle

\begin{abstract}
Uniswap v3 is the largest decentralized exchange for digital currencies, and 
allows a liquidity provider (LP) to allocate liquidity to one or more closed intervals of the price of an asset. An LP earns fee rewards proportional to the amount of its liquidity allocation when prices move in this interval. This induces the problem of {\em strategic liquidity provision}: smaller intervals result in higher concentration of liquidity and correspondingly larger fees when the price remains in the interval, but with higher risk as prices may exit the interval leaving the LP with no fee rewards. Although reallocating liquidity to new intervals can mitigate this loss, it comes at a cost, as LPs must expend gas fees to do so. We formalize the {\em dynamic liquidity provision problem,} and focus on a general class of strategies for which we provide a neural network-based optimization framework for maximizing LP earnings. We model a single LP  that  faces an exogenous sequence of price changes that arise from arbitrage and non-arbitrage trades in the decentralized exchange. We present experimental results informed by historical price data that demonstrate large improvements in LP earnings over existing allocation strategy baselines. Moreover we provide insight into qualitative differences in optimal LP behavior in different economic environments.
\end{abstract}

\section{Introduction}\label{sec:intro}
Decentralized finance (DeFi) is a large and rapidly growing collection of projects in the cryptocurrency and blockchain ecosystem. From May 2019 to May 2023, the TVL ({\em total value locked}, meaning the sum of all liquidity provided to the protocol) into DeFi protocols has increased 100x from 500 million USD to 50 billion USD. During this time period, TVL rapidly increased in 2021, reaching a peak of roughly 176 billion USD in November 2021, but it also suffered large drops in 2022 leading to current TVL levels.\footnote{\url{https://defillama.com/}} 

DeFi aims to provide the function of financial intermediaries and instruments through smart contracts executed on blockchains (typically Ethereum). The importance of decentralized exchanges (DEXes) is that traders can swap tokens of different types without a trusted intermediary. Most DEXes, including Uniswap, fall into the category of {\em constant function market makers} (CFMMs). Instead of using an order book as is done in traditional exchanges, CFMMs are smart contracts that use an {\em automated market  maker} (AMM)  to determine the price of a trade. 

In {\em Uniswap v2}, token pairs can be swapped for each other using {\em liquidity pools}, which contain quantities of each of the pair of tokens (say, token $A$ and token $B$).
 {\em  Uniswap v3}, which  launched on   Ethereum on May 3, 2021, introduced  
\textit{concentrated liquidity} \cite{adams2021uniswap}, where an LP can provide liquidity to each of any number of price intervals, these called {\em positions}.

As a result, Uniswap v3 supports a diversity of LP strategies in regard to the allocation of liquidity, as an LP can mint multiple positions, each on a different interval. Each LP is presented with a tradeoff between choosing  large positions that cover many possible prices but earn less fees and smaller more concentrated positions that are  more risky (since they cover fewer prices). Additionally, an LP can reallocate its liquidity as prices change, but this comes at two costs. First, an LP may need to trade between assets to mint new liquidity positions in a reallocation, potentially suffering losses from slippage in such trades. Second, since liquidity allocations are transactions that must be written as updates to the contract and included in a block, they also incur gas fees. Both of these costs must be incorporated in more complex LP strategies that make use of liquidity reallocation.  

Given the increased complexity for LPs in v3, it is important to understand the ways LPs can benefit from strategically allocating liquidity as prices change, as this ultimately impacts the performance of v3 contracts as DEXs. 
This paper develops optimal liquidity provision strategies, that  make use of liquidity reallocations and the full complexity of v3 positions. As in \cite{fan2022differential}, we adopt the perspective of an LP with stochastic beliefs over how market prices evolve, and how contract prices may change along with market prices via arbitrage and non-arbitrage trades. We also  make the same assumption as there, namely that the LP is small enough that these beliefs are independent of capital allocated by the LP. 
Given this, we provide a new
optimization framework for computing optimal dynamic liquidity provision strategies over a given time horizon, and  show that such strategies can provide large gains to LPs over strategies that are static or dynamic but with  simple liquidity positions.

 \medskip
 
{\bf Contributions.} We define {\it dynamic liquidity provision strategies}, as a means by which LPs can mitigate potential losses by using earned capital to reallocate liquidity to different price intervals in a given time horizon. In particular, we focus on the family of {\em $\tau$-reset strategies}, which allocate over an interval of prices centered on the current price and  whose width is controlled by $\tau$ (including the possibility of declining to allocate  liquidity) and reallocate  whenever the price moves outside this interval. Moreover, within the space of  dynamic liquidity provision strategies, we distinguish between those that are {\em context-aware} and {\em context-independent,} which depends on whether they incorporate historical price and contract information in their decision-making or not.

We develop the use of stochastic optimization for  optimizing over the space of $\tau$-reset strategies, considering an LP with stochastic beliefs on market and contract prices, and with  different levels of risk-aversion. We give empirical results based on historical Ethereum price data which show that our dynamic allocation strategies (which forcibly incorporate LP beliefs on price changes) give rise to large gains in LP earnings relative to baseline uniform allocation strategies (similar in nature to liquidity allocation in Uniswap v2). As expected, we see the best performance for context-aware allocation strategies, for which we make use of a neural network representation of an investment strategy. In addition, our results provide insight into how optimal LP behavior varies depending on relevant aspects of the economic environment. In particular, we find that more risk-averse LPs spread their liquidity over larger price ranges, especially when faced with a larger volume of non-arbitrage trades. In addition we also find that, as expected, optimal reset frequencies are sensitive to the reallocation costs.

\subsection{Related work}

The Uniswap v1 protocol was defined by \cite{uniswapv1white}, followed up with v2 by \cite{adams2020uniswap}, and most recently amended in v3 by \cite{adams2021uniswap}. There has been a growing body of work studying LP incentives in Uniswap v2. \cite{angeris2019analysis} present an analysis of Uniswap, and more broadly of constant-product AMMs, and demonstrate conditions for which the markets closely track the price on an external reference market. \cite{angeris2020improved} extend this line of research by demonstrating that the more general class of CFMMs incentivize participants to report the true price of an asset on an external market, demonstrating their value as price oracles.

\cite{aoyagi2020lazy} study the equilibrium liquidity provision of constant product AMMs, and show that strategic LPs  in the Uniswap v2 environment may have a non-monotonic best response when parameterized with the opponents' liquidity provision. \cite{angeris2020does} extend this line of work to  CFMMs and calculate bounds on the LP rewards based on the curve that defines the CFMM. \cite{capponi2021adoption} studies the adopation of decentralized exchanges more generally, using a sequential game framework to model interactions between LPs and traders. In addition, \cite{schlegel2022axioms} and \cite{frongillo2023axiomatic} provide an axiomatic framework for general CFMMs similar in nature to Uniswap v2, with the latter focusing on connections with CFMMs in the prediction market setting. 
\cite{evans2020liquidity} consider {\em  geometric mean market makers} (G3Ms), 
and show that 
passive liquidity provision can be used to replicate payoffs of financial derivatives and more active trading strategies. \cite{tassy2020growth} and \cite{uniswapsfinancialalchemy} analyze the growth in wealth of an LP in CFMM for a geometric Brownian motion price process. \cite{evans2021optimal} extend this to more general LP objectives and diffusion processes.

All above work applies to Uniswap v2 and similarly structured CFMMs but not to  v3. An early blog post by \cite{charmalphavault} describes a ``passive rebalancing'' strategy for v3, which aims at maintaining a 50-50 ratio of value for the assets of the  LP. 
 Most relevant to the present work, and also studying Uniswap v3, are the two papers, \cite{cartea2022decentralised, fan2022differential}, the first of which provides insight into how LPs can profit from liquidity reallocations with simple positions, and the second of which focuses on how LPs can profit from complicated static liquidity positions over a given time horizon.\footnote{Both of these papers were written after the first version of the present paper \cite{neuder2021strategic}} In more detail, the authors of \cite{cartea2022decentralised} provide a closed form solution for computing optimal LP allocations that dynamically readjust positions to different intervals as prices evolve over time. 
Though these strategies make use of dynamic reallocation of liquidity, the only allocations explored in the optimization are individual v3 positions over a single price range (which forcibly change at each reallocation) instead of the full class of potential allocations available to an LP in Uniswap v3. The authors of \cite{fan2022differential} compute optimal arbitrary v3 positions for small LPs who seek to maximize profit and loss over a fixed time horizon, 
but only consider static LP strategies which do not make use of reallocations as prices change.

Further related work on  v3 includes \cite{milionis2022automated}, who decompose divergence/impermanent loss into hedgeable market risk and profit made by arbitrage traders at a loss to the exchange. (This is related to \cite{cartea2022decentralised}, who decompose divergence/impermanent loss into two components: the loss due to arbitrage (convexity cost) and the cost of locking capital). 
\cite{bar2023uniswap} uses regret-minimization  from online learning to provide liquidity provision strategies under adversarial trading. \cite{milionis2023myersonian} and \cite{goyal2022finding}  study the construction of optimal CFMMs frwom the perspective of LP beliefs, with the latter providing a Myersonian framework for creating incentive compatible AMMs, and the former  employing techniques from convex optimization to determine optimal trading functions based on LP beliefs on future trades. \cite{heimbach2022risks} study the risks inherent to LP returns for multiple fixed strategies in different economic environments, concluding that liquidity provision in v3 is a sophisticated game where uninformed retail traders can suffer large disadvantages relative to more informed agents. In addition to studying optimal static liquidity provision, \cite{fan2022differential}   provide insights on how aspects of a v3 contract, notably the partition of price space, have implications on LP profit as well as  gas fees incurred by traders.

\subsection{Background on Uniswap v2 and v3}

Permitted trades in Uniswap v2 are determined by the {\em reserve curve}, $x y= L^2$, where $x$ and $y$ denote the the number of tokens of type $A$ and $B$ respectively in the liquidity pool, and the value of $L$ must be maintained across a trade. {\em Liquidity providers (LPs)} add tokens to liquidity pools for the traders to swap against, and are rewarded through the fees traders pay. 
An illustrative reserve curve is shown in Figure~\ref{fig:v2-reserve-trade}. Assuming that the v2 contract holds $x$ units of token $A$ and $y$ units of token $B$ with $xy = L^2$, then in order to sell some quantity $\Delta x>0$ of token $A$ for some quantity $\Delta y>0$  of token $B$, the trader must keep the product of reserves constant, with $\Delta y$ such that $(x-\Delta x) (y + \Delta y) = L^2$. 
This defines an effective contract price for token $A$ in units of token $B$, i.e., $P = -dy/dx$. In the context of the $xy = L^2$ curve, we have $y=L^2/x \implies P = L^2/x^2 = y/x$. By convention, we take the contract price to be the price of token $A$, which we may assume is volatile relative to token $B$.
In Uniswap v2, traders pay LPs a fee of 0.3\% of the transaction amount in return for using  the liquidity to execute a swap~\cite{adams2020uniswap}. In v2, the liquidity of every LP can be used for swaps at every possible price $P \in (0, \infty)$, and an LP earns fees based on their fraction of the total liquidity in the pool.
\begin{figure}[t]
\centering
\begin{tikzpicture}[scale=1.5]
  \draw[-{Straight Barb[scale=1.5]}, line width=1.2pt] (-0.5, 0) -- (3, 0) node[] {};
  \draw[-{Straight Barb[scale=1.5]}, line width=1.2pt] (0, -0.5) -- (0, 3) node[] {};

  \draw[blue!40, domain=1/3:3, smooth, samples=100, line width=1.7pt] plot (\x, {1/\x});

  \filldraw[black, radius = 0.7pt] (2/3, 3/2) circle node[below left] {$(x,y)$};
  \filldraw[black, radius = 0.7pt] (3/2, 2/3) circle node[below left] {$(x',y')$};

  \draw[dotted, latex-latex] (2/3, 3/2) -- (3/2, 3/2) node[midway, above] {$\Delta x$};
  \draw[dotted, latex-latex] (3/2, 3/2) -- (3/2, 2/3) node[midway, right] {$\Delta y$};

  \node[below] at (1.5,-0.1) {{\tt Token A reserves} ($x$)};

  \node[left, rotate=90] at (-0.25,2.5) {{\tt Token B reserves} ($y$)};

\end{tikzpicture}
\caption{The reserve curve for  Uniswap v2. If the pool has reserves $(x,y)$ where $x$ and $y$ represent units of token $A$ and $B$ respectively, then the contract price of token $A$ is $P = y/x$. A trader can send $\Delta x$ units of token $A$ to receive $\Delta y$ units of token $B$, such that $x'y' = L^2$, where $x' = x+\Delta x$ and $y' = y - \Delta y$. The  contract price of token $A$ after the trade is  $P' = y'/x'$.
\label{fig:v2-reserve-trade}}
\end{figure}

Considering Uniswap v3, the liquidity allocated by an LP to position $[P_a,P_b]$ earns  fees when the contract price is in that interval. If multiple LPs  allocate liquidity over an interval containing the current price, each is rewarded proportionally to the fraction of the liquidity they own at that price.
Figure~\ref{fig:v2-v3-affine} visualizes the functional invariant respected by the overall assets provided by LPs to support trades over $[P_a,P_b]$. For trades in this interval, a v3 contract effectively shifts the reserve curve of Uniswap v2  via an affine transformation to intercept the axes at $a$ and $b$, which  depend on the end points of interval $[P_a,P_b]$. This shifted curve is governed by the equation: $\left(x + L/\sqrt{P_b} \right) \left(y +L\sqrt{P_a}\right) = L^2$, and the intercepts, $a$ and $b$, can be calculated by letting $x$ or $y$ equal zero respectively~\cite{adams2021uniswap} .
This description of Uniswap v3 is inherently local, as it describes trade dynamics for a specific price interval. Gluing the local dynamics together for all prices gives rise to an {\em aggregate reserve curve} that governs arbitrary trades across all possible prices. This global reserve curve is in turn a function of the aggregate distribution of liquidity provided by all LPs. From the perspective of traders, when  more liquidity is allocated to a given price interval, there is less trade slippage for prices in that interval. This reduced slippage corresponds to a flatter section of the aggregate reserve curve, as visualized in Figure~\ref{fig:aggregate-visualization}.
\begin{figure}[t]
\centering
\begin{tikzpicture}[scale=1.5]
  \draw[-{Straight Barb[scale=1.5]}, line width=1.2pt] (-0.5, 0) -- (3, 0) node[right] {};
  \draw[-{Straight Barb[scale=1.5]}, line width=1.2pt] (0, -0.5) -- (0, 3) node[above] {};

  \draw[blue!40, domain=1/3:3, smooth, samples=100, line width=1.7pt] plot (\x, {1/\x});
  \draw[red, domain=0:5/6, smooth, samples=100, line width=1.7pt] plot (\x, {1/(\x+(2/3)) -(2/3)});

  \filldraw[black, radius = 0.7pt] (2/3, 3/2) circle node[above right] {$P_b$};
  \filldraw[black, radius = 0.7pt] (3/2, 2/3) circle node[above right] {$P_a$};
  \filldraw[black, radius = 0.7pt] (0, 5/6) circle node[above right] {$b$};
  \filldraw[black, radius = 0.7pt] (5/6, 0) circle node[above right] {$a$};

  \draw[->, line width=0.7pt] (1/2,4/3) -- (2/6, 7/6) node[] {};
  \draw[->, line width=0.7pt] (4/3,1/2) -- (7/6, 2/6) node[] {};

  \node[below] at (1.5,-0.1) {{\tt Token A reserves} ($x$)};

  \node[left, rotate=90] at (-0.25,2.5) {{\tt Token B reserves} ($y$)};

\node[draw, fill=white, line width = 1.2pt, rounded corners] at (2,2) (legend) {
  \begin{tabular}{@{}l@{}}
        \rule{0pt}{1.5ex}\tikz{\draw[blue!40, line width=1.5pt] (0,0) -- (20pt,0);} {\tt v2} \\
    \rule{0pt}{1.5ex}\tikz{\draw[red, line width=1.5pt] (0,0) -- (20pt,0);} {\tt v3}
  \end{tabular}
};

\end{tikzpicture}
\caption{The reserve curve of Uniswap v2 over all prices, and of Uniswap v3 over the price interval $[P_a,P_b]$. When trades give rise to contract prices in this interval, LP assets are swapped according to the v3 curve, which is an affine transformation of the v2 curve and defined to respect the price limits of  interval $[P_a,P_b]$.
\label{fig:v2-v3-affine}}
\end{figure}


\begin{figure}
  \begin{minipage}{0.4\textwidth}
    \centering
    \vspace{0.55cm}
        \scalebox{0.8}{%
        \begin{tikzpicture}
          \begin{axis}[
            width=10cm,
            height=6cm,
            axis lines=left,
            xmin=-3.5,
            xmax=3.5,
            ymin=0,
            ymax=0.45,
            xlabel=Price,
            ylabel=Aggregate Liquidity,
            xtick={-3,-2,-1,0,1,2,3},
            ytick={0,0.1,0.2,0.3,0.4},
            ticklabel style={font=\footnotesize},
            grid=none,
            samples=100,
            yticklabels={},
            ytick style={draw=none},
            xticklabels={},
            xtick style={draw=none},
          ]
            \addplot [ybar interval, fill=blue!50, draw=black, bar width=0.3]
              coordinates {(-3,0.004) (-2.7,0.013) (-2.4,0.033) (-2.1,0.063) (-1.8,0.101) (-1.5,0.148) (-1.2,0.199) (-0.9,0.242) (-0.6,0.272) (-0.3,0.287) (0,0.289) (0.3,0.277) (0.6,0.254) (0.9,0.222) (1.2,0.186) (1.5,0.149) (1.8,0.116) (2.1,0.088) (2.4,0.066) (2.7,0.049) (3,0.036)};
          \end{axis}
        
          \node[font=\small] at (0.1, -0.2) {0};
          \node[font=\small] at (4.2, -0.2) {1};
          \node[font=\small] at (8.3, -0.2) {$\infty$};
        \end{tikzpicture}
        }
    \label{fig:first-aggregate}
  \end{minipage}%
  \hspace{2.5cm}
  \begin{minipage}{0.4\textwidth}
    \centering
        \scalebox{0.7}{%
        \begin{tikzpicture}[scale=1.7]
          \draw[-{Straight Barb[scale=1.5]}, line width=1.2pt] (0, 0) -- (3, 0) node[] {};
          \draw[-{Straight Barb[scale=1.5]}, line width=1.2pt] (0, 0) -- (0, 3) node[] {};
        
          \draw[dotted, blue!40, domain=1/3:3, smooth, samples=100, line width=1.7pt] plot (\x, {1/\x});
          \draw [red, line width=1.7pt] (0.2,3) .. controls (0.3,2.5) and (0.1,2.2) .. (1.2,1.2) .. controls (2,0.5) and (2.5,0.2)  .. (3,0.2);
        
          \node[below] at (1.5,-0.1) {{\tt Token A reserves} ($x$)};
        
          \node[left, rotate=90] at (-0.25,2.5) {{\tt Token B reserves} ($y$)};
        
        \node[draw, fill=white, line width = 1.2pt, rounded corners] at (2.5,2) (legend) {
          \begin{tabular}{@{}l@{}}
                \rule{0pt}{1.5ex}\tikz{\draw[dotted, blue!40, line width=1.5pt] (0,0) -- (20pt,0);} {\tt v2} \\
            \rule{0pt}{1.5ex}\tikz{\draw[red, line width=1.5pt] (0,0) -- (20pt,0);} {\tt v3}
          \end{tabular}
        };
        \end{tikzpicture}
        }
    \label{fig:second-aggregate}
  \end{minipage}
\vspace{-0.5cm}
\caption{An aggregate distribution of liquidity for a Uniswap v3 contract (left plot), with most liquidity  allocated close to unit price $P=1$. This   results in an aggregate reserve curve (right, red line), which is flatter than the corresponding v2 curve (dotted blue) at prices close to $P=1$ and supports a larger volume of trades at these prices with less slippage.
\label{fig:aggregate-visualization}}
\end{figure}

\subsection{Outline}

Section~\ref{sec:uniswap-mechanics}  introduces the Uniswap v3 protocol. Section~\ref{sec:liq-allocations-LP-earnings} formalizes the earnings to an LP from a dynamic liquidity provision strategy and introduces the family of $\tau$-reset strategies. Section~\ref{sec:optimization} introduces the computational methods for optimizing over $\tau$-reset strategies and specifically defines the context-aware/independent dynamic liquidity strategies we empirically study as well as simple baselines to which we compare performance. Section~\ref{sec:experimental-setup} provides details regarding the economic environments we modulate to study optimal LP behavior, and Section~\ref{sec:results} presents empirical results. Section~\ref{sec:conclusion-new} gives open problems for future research and concludes.

\section{The Mechanics of Uniswap}
\label{sec:uniswap-mechanics}

In this section, we provide a brief overview of Uniswap v3 contracts. In all that follows, we consider a v3 contract that enables trades between two types of tokens that we designate token $A$ and token $B$. Furthermore, without loss of generality, we assume  token $B$ is the numeraire, hence when we refer to the price in the contract, we  refer to the price of a unit of  $A$ in terms of  $B$. As mentioned in the introduction,  liquidity providers (LPs) provide bundles of $A$ and $B$ tokens to the contract as liquidity to be traded against. The following sections largely follow the mathematical formalism of \cite{fan2022differential}, to which we refer the reader for more in-depth details regarding Uniswap v2 and v3 contract dynamics.

\subsection{v3 Contracts}

\paragraph*{Partitioned Price Space} 
In Uniswap v2 contracts, the liquidity that LPs provide is used to support every trade, whatever the trade price. Uniswap v3 contracts provide a richer set of actions for an LP, where they can specify a price range where  liquidity is to be used for trading. In order to enable this functionality, a v3 contract partitions token $A$ prices into a finite set of {\it price buckets}: $\vect{\mu} = \{B_{-m},\dots,B_0,\dots,B_n\}$ with buckets $B_i = [a_i,b_i]$. We also require $a_0 < 1 < b_0$, so that the parity price lies in the $0$-th bucket,
 and that $b_i = a_{i+1}$ for $i \in \{-m, \dots, n-1\}$. 

\paragraph*{Contract Price}
A v3 contract maintains the {\it contract price} of token $A$,  $P \in (0,\infty)$.  This
is the infinitesimal price that traders obtain for trading with the contract.

\paragraph*{Minting and Burning Liquidity}

LPs provide (or {\em mint}) liquidity in a particular bucket, $B_i=[a_i,b_i]$, referred to as {\em  $B_i$-liquidity}, by sending  a  bundle of $A$ and $B$ tokens  to the contract. The  token bundle required to mint $L$ units of $B_i$-liquidity is given by the {\em  liquidity value function}~\cite{fan2022differential}, and is tuple $\cV^{(3)}(L,P,B_i)$, where the first component is the quantity in token $A$ and the second is the quantity in token $B$.
For $a < b$,  let $\Delta^x_{b,a} = \frac{1}{\sqrt{a}} - \frac{1}{\sqrt{b}}$ and $\Delta^y_{a,b} = \sqrt{b} - \sqrt{a}$. 
\begin{definition}[$B_i$-Liquidity Value~\cite{fan2022differential}]
\label{def:value-function}
For contract price $P$, bucket $B_i=[a_i,b_i]$,  and number of units $L>0$,
the {\em bundle liquidity value} $\cV^{(3)}(L,P,B_i) \in \mathbb{R}^+ \times \mathbb{R}^+$
is defined as
\begin{equation}
\mathcal{V}^{(3)}(L,P,B_i) = \left\{
        \begin{array}{ll}
            (L \Delta^x_{b_i,a_i}, 0)& \quad \mbox{if $P < a_i$,} \\
            (0,L \Delta^y_{a_i,b_i}) & \quad \mbox{if $P > b_i$, or} \\
            (L\Delta^x_{b_i,P}, L\Delta^y_{a_i,P}) & \quad \mbox{if $P\in B_i$,} 
        \end{array}
    \right.
\end{equation}
and specifies the bundle of $A$ and $B$ tokens, respectively, 
 to mint $L$ units of $B_i$-liquidity.
\end{definition}
LPs can also remove liquidity they have a claim to from the contract by means of the value function. When this happens, we say that an LP {\it burns} $L$ units of $B_i$ liquidity, and the LP receives  token bundle $\cV^{(3)}(L,P,B_i)$ if the contract price is $P$. Since $\cV^{(3)}$ is linear in $L$,  we also adopt shorthand $\cV^{(3)}(P,B_i) = \cV^{(3)}(1,P,B_i)$ for the token bundle value of a single unit of liquidity, with $\cV^{(3)}(L,P,B_i) = L \cdot \cV^{(3)}(P,B_i)$.   

\paragraph*{Contract State}
The contract price $P$ and the collective set of minted liquidity allocations of all LPs denotes the
{\em  state} of a v3 contract. As shown in \cite{fan2022differential} (Section 2.2.3) we can simplify the set of potential states the contract can take by making two minor assumptions regarding LPs:
\begin{enumerate}
    \item Every LP allocates liquidity to a single bucket from $\vect{\mu}$, and
    \item Every bucket from $\vect{\mu}$ has liquidity allocated by at least one LP.  
\end{enumerate}

In regard to the first assumption, Uniswap v3 contracts actually allow an LP to mint positions over multiple contiguous buckets, such as minting a position worth $L$ units of liquidity over the price interval $[a_i,b_j]$ where $i < j$. However, this  is equivalent to minting positions worth $L$ units of liquidity for each bucket in $\{B_i,\dots,B_j\}$. 
With regards to the second assumption, it can indeed be the case that no LP allocates liquidity to a given bucket, however in practice contract prices do not reach buckets without liquidity. Indeed most liquidity is allocated to multiple buckets around contract prices.\footnote{\url{https://info.uniswap.org/home\#/pools/}} Moreover, trade dynamics when prices reach buckets with no liquidity can be approximated by the dynamics that occur when the bucket has an infinitesimal amount of liquidity. 

Given the assumptions, at any given moment there are (a variable) number, $d \geq n+m+1$, of LPs providing liquidity to the contract. Furthermore, we let $\sigma: [d] \rightarrow \{-m,\dots,0,\dots,n\}$ and $\vect{L} = (L_1,\dots,L_d)$ be such that the $j$-th LP has minted $L_j$ units of $B_{\sigma(j)}$-liquidity. We call $(\sigma,\vect{L})$ the {\em allocation profile} and use this to define the contract state.  
\begin{definition}[Contract state]
Given contract price $P$ and allocation profile $(\sigma,\vect{L})$,  the {\em state} of a v3 contract is  $S = (P,\sigma,\vect{L})$. 
\end{definition}

As we have seen above, LPs can change the contract state by minting and burning liquidity. Traders  alter the contract state by changing the contract price $P$. 

\paragraph*{Trading and Fees}

Suppose $P$ does not lie on the boundary of any bucket, and refer to the bucket in which it is contained as the {\it active bucket}, with index $i^* \in \{-m,\dots,n\}$. The LPs who have minted liquidity in $B_{i^*}$ are the {\it active LPs}, and the sum of their   liquidity  in this bucket is the {\it active liquidity},  $L^*  = \sum_{j \in \sigma^{-1}(i^*)} L_j$. Let $\cV^{(3)}(L^*,P,B_{i^*}) = (x^*,y^*)$ denote the {\it active token bundle}. Let $\gamma \in (0,1)$ denote the {\it fee rate} of the contract, which specifies what portion of trades are skimmed as fees for LPs (Uniswap v3 contracts have three potential fee rates, $\gamma \in \{0.0005, 0.003, 0.01\}$). 
Since $P$ lies in the interior of $B_{i^*}$, we have $a_i^* < P < b_i^*$. We let $\bar{x} = (\cV^{(3)}(L^*,b_{i^*},B_{i^*}))_1 = L \Delta^x_{b_i^*,a_i^*}$, and as shown in \cite{fan2022differential} (Section 2.2.3), the fact that $(x^*,y^*)$ is the active token bundle for a price in the interior of the active bucket implies that $x^* < \bar{x}$, as visualized in Figure~\ref{fig:v3-trade-dynamics}. 

\begin{figure}
\centering
\begin{tikzpicture}[scale=1.7]
  \draw[-{Straight Barb[scale=1.5]}, line width=1.2pt] (-0.5, 0) -- (3, 0) node[right] {};
  \draw[-{Straight Barb[scale=1.5]}, line width=1.2pt] (0, -0.5) -- (0, 3) node[above] {};

  \draw[blue!40, domain=1/3:3, smooth, samples=100, line width=1.7pt] plot (\x, {1/\x});
  \draw[red, domain=0:5/6, smooth, samples=100, line width=1.7pt] plot (\x, {1/(\x+(2/3)) -(2/3)});

  \filldraw[black, radius = 0.7pt] (2/3, 3/2) circle node[above right] {$b_i^*$};
  \filldraw[black, radius = 0.7pt] (3/2, 2/3) circle node[above right] {$a_i^*$};
  \filldraw[black, radius = 0.7pt] (1, 1) circle node[above right] {$P$};
  \filldraw[black, radius = 0.7pt] (0, 5/6) circle node[left] {$\bar{y}$};
  \filldraw[black, radius = 0.7pt] (1/3, 1/3) circle node[above right] {};
  \filldraw[black, radius = 0.7pt] (5/6, 0) circle node[below] {$\bar{x}$};

  \draw[dotted] (1/3, 1/3) -- (0, 1/3) node[left] {$y^*$};
  \draw[dotted] (1/3, 1/3) -- (1/3, 0) node[below] {$x^*$};

  \draw[->, line width=0.7pt] (1/2,4/3) -- (2/6, 7/6) node[] {};
  \draw[->, line width=0.7pt] (4/3,1/2) -- (7/6, 2/6) node[] {};
  \draw[->, line width=0.7pt] (5/6,5/6) -- (2/3, 2/3) node[] {};

  \node[below] at (1.5,-0.25) {{\tt Token A reserves} ($x$)};

  \node[left, rotate=90] at (-0.5,2.5) {{\tt Token B reserves} ($y$)};

\node[draw, fill=white, line width = 1.2pt, rounded corners] at (2,2) (legend) {
  \begin{tabular}{@{}l@{}}
        \rule{0pt}{1.5ex}\tikz{\draw[blue!40, line width=1.5pt] (0,0) -- (20pt,0);} {\tt v2} \\
    \rule{0pt}{1.5ex}\tikz{\draw[red, line width=1.5pt] (0,0) -- (20pt,0);} {\tt v3}
  \end{tabular}
};

\end{tikzpicture}
\caption{Visualization for trade dynamics in an active bucket given by $B_i^* = [a_i^*,b_i^*]$. The blue curve is given by $xy = (L^*)^2$, where $L^*$ is the active liquidity in $B_i^*$. The contract price, $P$, corresponds to the active bundle $(x^*,y^*)$. The upper bound on the amount of token $A$ which can be present in an active bundle for $B_i^*$ is given by $\bar{x}$. Similarly, the upper bound on the amount of token $B$ which can be present in an active bundle for $B_i^*$ is given by $\bar{y}$.
\label{fig:v3-trade-dynamics}}
\end{figure}

Suppose  a trader wants to sell $\Delta x \leq \frac{1}{1-\gamma}(\bar{x} - x^*)$  units of token $A$ for some quantity of token $B$. The contract first  takes $\gamma \Delta x$ units of token $A$ as fees for LPs, to be shared proportionally amongst active LPs where an active LP $j$ with $\sigma(j) = i^*$  receives $\gamma \Delta x \frac{L_j}{L^*}$ units of token $A$.
The remaining $(1-\gamma) \Delta x \leq \bar{x} - x^*$ units of token $A$ are used to change the contract state and determine how many units of token $B$ the trader receives.
There is a unique $P' \in B_i$ such that $P' < P$ and $\cV^{(3)}(L^*,P',B_i) = (x',y')$ with $x'= x^* + (1-\gamma)\Delta x$. $\Delta y = y^* - y'$ is the quantity of $B$ tokens  received in exchange for $\Delta x$ units of $A$ tokens and the price changes from $P$ to $P'$. 
For a trader who wants to sell  $\Delta x > \frac{1}{1-\gamma}(\bar{x} - x^*)$ units of token $A$, then after skimming fees, $(1-\gamma)\Delta x > \bar{x}-x^*$ units of $A$ are used to change  the contract state by shifting the contract price. First, $\bar{x}-x^*$ is used to change the contract state, shifting the contract price to $P' = a_i$. This changes the active bundle  to $i^*-1$ (as the price decreases), and the remaining amount $(1-\gamma)\Delta x - (\bar{x} - x^*) > 0$ is traded recursively, as specified above.
This works symmetrically for a trader who wants to sell token $B$, with the main difference that these trades increase the contract price.  Also, if the initial price before a trade is not in the interior of a bucket, but rather $P \in B_i \cap B_{i+1}$ then depending on whether the trader sells token $A$ (decreasing the contract price) or sells token $B$  (increasing the contract price), the active bucket is  $i$ or $i+1$ respectively.

\section{Liquidity Allocation Strategies and LP Earnings}
\label{sec:liq-allocations-LP-earnings}

In this section, we describe a rich set of strategies that LPs can use to maximize their earnings over a given time horizon. As token $B$ is the numeraire, we measure all earnings in terms of units of token $B$ and we assume  that the LP begins with a fixed budget consisting of $W > 0$ units of token $B$. We model price discovery between $A$ and $B$ tokens as occurring outside of Uniswap contracts, and in addition to the contract price there is  a {\it market price} that is determined by external markets. We denote the market price by $P_m$ and contract price by $P_c$, and we extend price $P$ so that $P = (P_m,P_c)$ denotes a {\it contract-market price pair}.
Furthermore, we assume that arbitrage trade can be performed by  traders at price $P_m$ which in turn brings $P_c$ close to $P_m$ (see Section~\ref{sec:Modeling-Contract-Market-Prices}).
In what follows, we consider time horizons that are characterized by a single sequence of $T>0$ contract-market prices, denoted by $\vect{P} = (P_0,\dots,P_T)$, where $P_t = (P_{c,t},P_{m,t})$ is the $t$-th contract-market price in the sequence (time steps are indexed $t$).

\subsection{Static Liquidity Provision Strategies}
\label{sec:static-LP-earnings}

We begin by using a similar mathematical formalism and notation from \cite{fan2022differential} to express an LP's earnings over price sequence $\vect{P}$ for a simple family of liquidity allocation strategies. 

\begin{definition}[Static liquidity provision strategy] 
An LP with an initial budget of $W>0$ units of token $B$ uses a {\em static liquidity provision strategy}
when they
\begin{enumerate}
\item  mint an initial liquidity allocation at $P_0$, 
\item  accrue token fees over the course of $\vect{P}$ as prices change, and
\item burn the existing liquidity allocation at $P_T$, the end of the time horizon, to recover invested capital from the contract. 
\end{enumerate}
\end{definition}

We focus on a single LP, and suppose  they mint liquidity positions at the beginning of $\vect{P}$, when contract-market prices are given by $P_0 = (P_{c,0}, P_{m,0})$, with their initial budget of $W>0$ units of token $B$. For this, let  $\vect{x} = (x_{-m},\dots,x_n)$ denote a {\it proportional liquidity allocation}, where for $i \in \{-m,\dots,n\}$, $x_i\geq 0$ represents the proportion of capital used to mint $B_i$-liquidity (with $\sum_{i=-m}^{n} x_i \leq 1$  so that  $\vect{x} \in \Delta^{m+n+2}$, the $(m+n+2)$-dimensional simplex). 
The LP uses $W x_i$ units of token $B$ to mint $B_i$-liquidity for each of $i \in \{-m,\dots,n\}$.  
Let $x_{n+1} = 1 - \sum_{i=-m}^n x_i \in [0,1]$ denote the proportion of capital that the LP does not invest and keeps as units of token $B$. 
 
Let $\cB: (\mathbb{R}^+)^2 \times \mathbb{R}^+ \rightarrow \mathbb{R}^+$, defined as $\cB((z_1,z_2),P_m) = P_m \cdot z_1 + z_2,$ return the {\em token $B$ market worth} of a bundle of $A$ and $B$ tokens when token $A$ has market price  $P_m$. 
For a given contract-market price sequence, $\vect{P}$, let $w_i = \cB(\cV^{(3)}(P_{c,0},B_i),P_{m,0})$ denote the amount of $B$ tokens required to mint one unit of $B_i$-liquidity at the initial contract-market price of $P_0 = (P_{c,0},P_{m,0})$. With this in hand,  let  vector $\vect{\ell} = (\ell_{-m},\dots,\ell_n)$ denote the {\it absolute liquidity allocation} induced by initial budget $(W)$, proportional liquidity allocation $(\vect{x})$ and initial contract-market price $(P_0)$. It follows that $\ell_i = \frac{W x_i} {w_i}$ units of $B_i$-liquidity for each $i \in \{-m,\dots,n\}$. This implies that $\vect{\ell}$ is linear as a function of each of $\vect{x}$ and $W$.  

\subsubsection{Linearity of Fee Rewards in \texorpdfstring{$\vect{x}$}{x}}

We are ultimately interested in expressing an LP's token $B$ value of  earnings as a function of their liquidity allocation over the contract-market price sequence. We begin by determining the amount of trading fees earned by an LP. Interestingly, these earnings are not only independent of other LP allocations, but also linear in $\vect{\ell}$ (and consequently $\vect{x}$). 
\begin{theorem}[Section 3.1 \cite{zhao2021understand}]\label{thm:indp}
For a fixed contract-market price sequence $\vect{P}$, the amount of $A$ tokens and $B$ tokens accrued from fees is linear in $\vect{\ell}$ and independent of the liquidity of other LPs in the contract. 
\end{theorem}

That the fees that a single LP earns are independent of other LPs' liquidity allocations follows from the assumption that contract-market prices are independent of the liquidity allocation of this LP. Indeed, for a fixed price sequence,  allocating liquidity by an LP has  two effects. First, increasing the liquidity means that a larger volume of trade needs to happen to effect the same price change, resulting in more fees to be paid out to LPs.  Second, the same LP has a proportionally larger amount of liquidity across the relevant price interval. The net effect is that fees are only a function of a single LP's proportional (or absolute) liquidity allocation. Theorem~\ref{thm:indp} justifies the following definition of a trading fee function for a single LP and a given contract-market price sequence.
\begin{definition}[Trading Fee Functions]
Suppose that $\vect{P}$ is a fixed contract-market price sequence and  $W>0$  an initial token $B$ budget. For a proportional (or absolute) liquidity allocation given by $\vect{x}$ (or $\vect{\ell}$), we let $F^A(\vect{x},W,\vect{P})$ (or $F^A(\vect{\ell},\vect{P})$) denote the units of $A$ tokens earned from fees over $\vect{P}$ from downward price movements. Similarly, we let $F^B(\vect{x},W,\vect{P})$ (or $F^B(\vect{\ell},\vect{P})$) denote the units of $B$ tokens earned from fees over $\vect{P}$ from upward price movements.
\end{definition}

When the resulting absolute liquidity allocation $\vect{\ell}$ is treated as a function of $\vect{x}$ and $W$, it is linear in $\vect{x}$ and $W$, and it follows that both $F^A(\vect{x},W,\vect{P})$ and $F^B(\vect{x},W,\vect{P})$ are linear in $\vect{x}$ and $W$. For this reason, we let $F^A(\vect{x},\vect{P}) = F^A(\vect{x},1,\vect{P})$ and $F^B(\vect{x},\vect{P}) = F^B(\vect{x},1,\vect{P})$. This in turn implies that $F^A(\vect{x},W,\vect{P}) = W \cdot F^A(\vect{x},\vect{P})$, and similarly $F^B(\vect{x},W,\vect{P}) = W \cdot F^B(\vect{x},\vect{P})$ for arbitrary $\vect{x},W,$ and $\vect{P}$.

\subsubsection{Burning Liquidity Allocations at \texorpdfstring{$P_T$}{P\_T}}

All that remains to fully quantify the earnings of an LP over $\vect{P}$ is to take into account the capital they obtain by burning their liquidity positions at time $T$ under contract-market price $P_T = (P_{c,T},P_{T,m})$, obtaining a final quantity of token $B$.
For this, let $w_i' = \cB(\cV^{(3)}(P_{c,T},B_i),P_{m,T})$ be the token $B$ worth of the capital obtained from burning 1 unit of $B_i$-liquidity at the final contract-market price of $P_T = (P_{c,T},P_{m,T})$. Given absolute liquidity position $\vect{\ell}$, the overall token $B$ value of capital obtained from burning is $C(\vect{\ell},\vect{P}) =  \sum_{i=-m}^n w_i' \ell_i$, and linear in $\vect{\ell}$. 
Let $C(\vect{x},W,\vect{P})$ denote the final token $B$ worth (at $P_T$) of a liquidity position minted at $P_0$ with $\vect{x}$ and budget $W$. Since proportional allocations also allow an LP to maintain funds in terms of token $B$ (i.e., when $x_{n+1} > 0)$, we obtain the expression $C(\vect{x},W,\vect{P}) = C(\vect{\ell},\vect{P}) + Wx_{n+1}$, where $\vect{\ell}$ is the  absolute liquidity allocation corresponding to $\vect{x}$.
Once more, since $\vect{\ell}$ is in turn linear in $\vect{x}$ and $W$, it follows that $C$ is  linear in $\vect{\ell}$ and $W$. For this reason, we let $C(\vect{x},\vect{P}) = C(\vect{x},1,\vect{P})$, so that $C(\vect{x},W,\vect{P}) = W \cdot C(\vect{x},\vect{P})$ for arbitrary $\vect{x}, W$, and $\vect{P}$.

\subsubsection{Linearity of Overall Earnings in \texorpdfstring{$\vect{x}$}{x}}

We now define an LP's earnings over a contract-market price sequence. 
\begin{definition}
Suppose that $\vect{P} = (P_0,\dots,P_T)$ is a contract-market price sequence and that $\vect{x} \in \Delta^{m+n+2}$ is a proportional liquidity allocation. An LP's earnings (in units of token $B$) under $\vect{P}$ with an initial budget of $W > 0$ is,
    $$
    V(\vect{x},W,\vect{P}) = P_{m,T} \cdot F^A(\vect{x},W,\vect{P}) + F^B(\vect{x},W,\vect{P}) + C(\vect{x},W,\vect{P}).
    $$
\end{definition}

From this definition, we conclude that  an LP's earnings from a fixed contract-market price sequence
and with a static liquidity provision strategy
are  linear in $\vect{x}$ and $W$.\footnote{As a side note, with the definition of LP's earnings here we are modeling the profit and loss (PnL) of an LP who holds their Uniswap v3 liquidity position without hedging. The strategy optimization approach in this paper can be naturally extended for maximizing the PnL of a delta-hedge LP as well by incorporating Loss versus Rebalancing (LVR) from \cite{milionis2022automated} since LVR is also linear in $\vect{x}$ and $W$.}

\begin{theorem}
$V$ is linear in both $W$ and $\vect{x}$ for any contract-market price sequence, $\vect{P}$. 
\end{theorem}

\begin{proof}
This is an immediate corollary of the fact that $F^A$, $F^B$ and $C$ are each linear in $\vect{x}$ and $W$ for any contract-market price sequence $\vect{P}$. 
\end{proof}

Similar to before, we use the shorthand $V(\vect{x},\vect{P}) = V(\vect{x},1,\vect{P})$ such that $V(\vect{x},W,\vect{P}) = W \cdot V(\vect{x},\vect{P})$ for arbitrary $\vect{x}, W$, and $\vect{P}$. 

\subsection{Dynamic Liquidity Provision Strategies}
\label{sec:dynamic-liquidity-provision}

In this section, we introduce the notion of {\it dynamic liquidity provision} strategies, 
where an LP can reallocate their liquidity at any time step of the contract-market price sequence, $\vect{P}$. 

At a high level, if an LP chooses to reallocate liquidity at time $t$, they burn their existing liquidity position and use their overall earnings at time $t$, denoted by $W_t$, to mint a new proportional allocation, $\vect{x}$, given the contract-market price $P_t$. Reallocation comes at a cost however, which represents the fact that burning and minting positions on a Uniswap contract requires paying gas fees, and that minting new positions may require the LP to trade between $A$ and $B$ tokens.
We model reallocation costs as proportional to overall earnings used to mint the position $\vect{x}$ (i.e. the funds corresponding to capital kept token $B$ outside the contract ($x_{n+1}$) do not incur a cost). Cost is specified by a single parameter, $\eta \in [0,1]$, such that the LP retains $\eta W_t (1 - x_{n+1})$ of the funds they intend to use for minting a new position after paying reallocation costs at time $t$ (i.e., by paying $(1-\eta) W_t (1 - x_{n+1})$ in reallocation cost).  

We partition price sequence $\vect{P}$ into {\it epochs}, which are sequences of contract-market prices from $\vect{P}$ uninterrupted by liquidity reallocations. For an LP that burns and reallocates liquidity positions at time steps $\vect{t} = (t^0,\dots t^k)$, where $t^0 = 0$ and $t^k = T$, there are $k \geq 1$ epochs, where the $j$-th epoch is  $E^j = (P_{t^j},\dots,P_{t^{j+1}})$. When indexing over epochs we use superscripts, and when indexing over time-steps in $\vect{P}$ we use subscripts. 

Each epoch, $E^j$, is   associated with the total earnings, $W^j$, the LP has accrued at the beginning of the epoch, and  the proportional liquidity allocation, $\vect{x}^j$, the LP uses to mint positions with $W^j$ over $E^j$. From the static liquidity allocation analysis, it follows that the LP's earnings over the epoch are given by $W^j \cdot V(\vect{x}^j,E^j)$. After incorporating the proportional reallocation cost,  the earnings available for the LP to use for $E^{j+1}$ are $W^{j+1} = \eta W^j \cdot V(\vect{x}^j,E^j)(1 - x^{j+1}_{n+1})$. We encode the collection of all proportional allocations as a $\left( k \times (m+n+1) \right)$ matrix $\matr{X}$, such that the $j$-th row of $\matr{X}$ corresponds to $\vect{x}^j$. 
 
\begin{definition}[Dynamic liquidity provision strategy]
We say that $\Lambda$ is a {\em  dynamic liquidity provision strategy}
 if it takes as input a contract-market price sequence, $\vect{P} = (P_1,\dots,P_T)$, and defines:
\begin{itemize}
	\item $\vect{t} = (t^0,\dots,t^k)$, with $k \geq 1$, $t^0 = 0$, and $t^k = T$. These are time-steps where a reallocation occurs.
	\item $\matr{X} \in mat\left( k \times (m+n+1) \right)$ such that the $j$-th row of $\matr{X}$ encodes $\vect{x}^j$, the proportional liquidity allocation profile to be used at $E^j$.
\end{itemize}
We write $\Lambda(\vect{P}) = (\vect{t},\matr{X})$ and say that this is the {\em realized dynamic liquidity provision strategy} of an LP under $\Lambda$ for contract-market price sequence $\vect{P}$.
\end{definition}

For an initial budget $W (= W_0 = W^0)$, we let $V(\Lambda,W,\vect{P})$ denote the overall earnings an LP obtains over $\vect{P}$ by employing strategy $\Lambda$, which can be computed recursively over the epochs of $\vect{P}$. As in previous sections, we let $V(\Lambda,\vect{P}) = V(\Lambda,1,\vect{P})$.

\subsubsection{Reset Liquidity Strategies}

In practice, a  strategy $\Lambda$ may not be implementable, for example requiring an LP to  know 
the full contract-market price sequence, $\vect{P}$, before it is realized. In this section, we focus on a specific family of implementable dynamic liquidity provision strategies, the {\it reset liquidity strategies}. 

For this, an LP at time-step $t$ with accumulated earnings  $W_t$ may choose to {\em  trigger} a liquidity reallocation based on the contract-market price sequence up to time $t$, denoted $\vect{P}_{\leq t}$. For reset liquidity strategies, the LP maintains a {\it reference bucket index} $Z_t \in \{-m,\dots,n\}$ (correspondingly a reference bucket $B_{Z_t} \in \vect{\mu}$). We let $S_t = (Z,W_t,\vect{P}_{\leq t})$ denote the {\em system  state}, and we let $\cS$ denote the space of all possible system states. A liquidity reset  consists in updating the reference bucket index and using the $W_t$ units of $B$ tokens at their disposal to mint a liquidity position relative to the reference bucket index $Z_t$. 
\begin{definition}[Reset liquidity provision strategy]
A {\em reset liquidity provision strategy (reset-LP strategy)}
 is composed of:
\begin{enumerate} 
    \item A {\em reference bucket update function}, $g$, which takes as input an arbitrary system state $S \in \cS$ and updates the reference bucket index to $Z \leftarrow Z'$ where $Z' = g(S)$. 
    \item An {\em allocation function}, $A: \cS \times \mathbb{Z} \rightarrow [0,1]$, which specifies the fraction of budget an LP allocates to mint liquidity in each bucket relative to $Z$ after a liquidity reset is triggered. More specifically, $A$ gives rise to the proportional allocation $\vect{x}$ such that $x_{i} = A(S,i-Z)$.
    \item A {\em reset condition},  $h(S) \in \{0,1\}$, which is
 an indicator function for whether a reset is triggered in system state $S\in \cS$
 and specifies which contract-market prices, relative to the reference bucket, will
trigger a liquidity reset. 
 In the event of a trigger, a new reference bucket is computed via update function $g$. 
\end{enumerate}
We  denote a reset-LP strategy by tuple $(g,h,A)$. 
\end{definition}

Of particular interest is the family of {\em $\tau$-reset strategies}. These strategies have LPs reset liquidity when the index of the bucket containing the contract price is more than $\tau$ away from the reference index, $Z$. In the case of a reset, the reference bucket changes to the bucket containing the current contract price.
\begin{definition}[$\tau$-reset Strategy]
Suppose that $\tau$ is a non-negative integer. We let $h_\tau: \cS \rightarrow \{0,1\}$ and $g_\tau: \cS \rightarrow \{-m,\dots,n\}$ denote trigger and reference bucket update functions, defined 
for system state $S_t = (Z_t,W_t,\vect{P}_{\leq t}) \in \cS$ as
\begin{itemize}
\item $h_\tau(Z_t,W_t,\vect{P}_{\leq t}) = 1$ if and only if $P_{c,t} \in B_i$ and $|Z_t - i| > \tau$, and
\item $g_\tau(Z_t,W_t,\vect{P}_{\leq t}) = P_{c,t}$.
\end{itemize}
We say that $(g_\tau,h_\tau,A)$ is a {\em $\tau$-reset strategy}, for any allocation function, $A:\cS \times \mathbb{Z} \rightarrow [0,1]$.
\end{definition}

\begin{figure}
    \centering
    \includegraphics[width=0.7\linewidth]{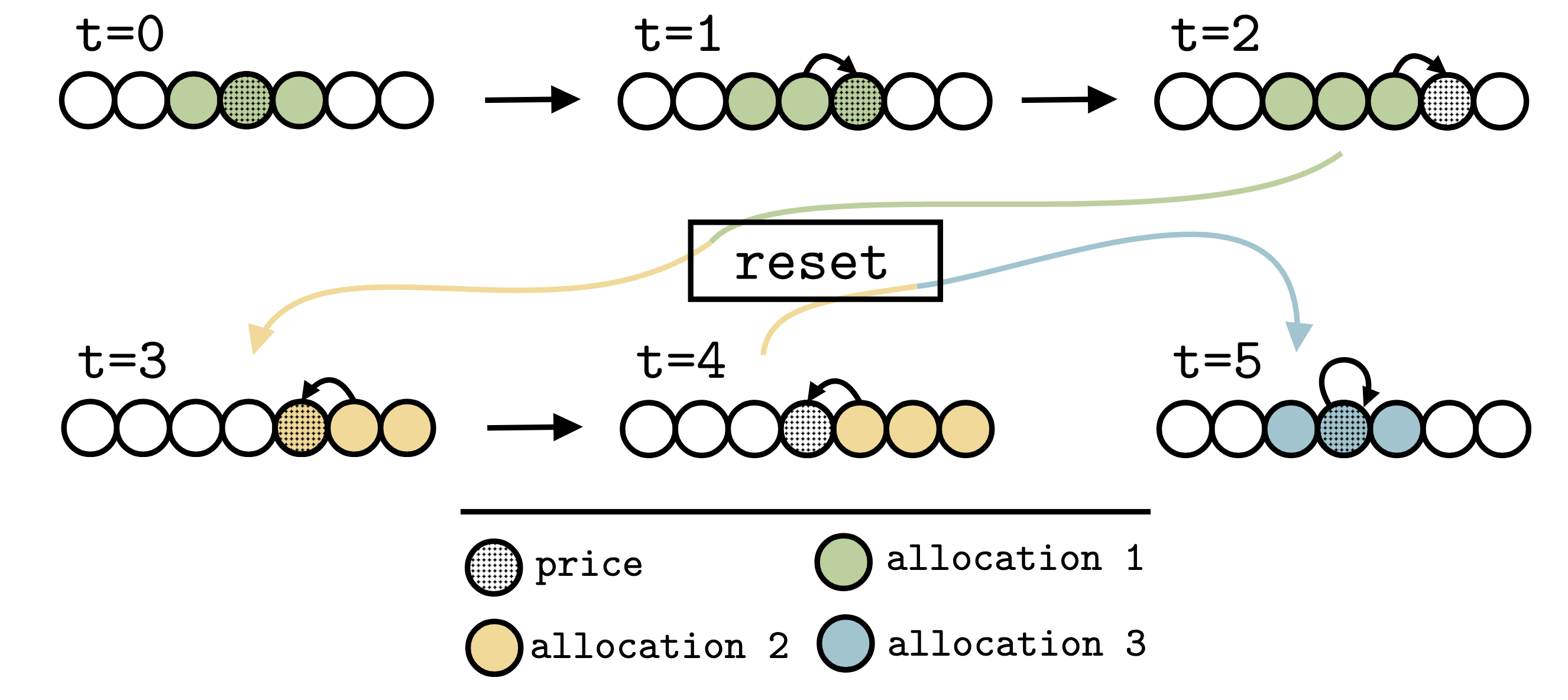}
    \caption{An illustration of how a $\tau$-reset strategy with $\tau = 1$ can play out. Buckets are represented by circles, and for simplicity we assume that market and contract prices move together at each time step. The shaded circle represents the bucket that contract/market prices are in, and the dynamics of price movements are expressed by the smaller arrows between buckets. Colored buckets represent the contiguous $2\tau + 1$ buckets centered around an epoch's reference bucket. For this sequence, we see that price movements at $t_1 = 2$ and $t_2 = 4$ trigger resets, as the shaded bucket escapes the contiguous $2\tau + 1$ colored buckets. The specific reallocation after each trigger is specified by the allocation function $A$  in the $\tau$-reset strategy. \label{fig:reset_strat}}
\end{figure}

We illustrate the versatility of $\tau$-reset strategies through some examples:
\begin{enumerate}
    \item (Static Strategies): For $\tau > T$, i.e., the time horizon of $\vect{P}$, we recover static strategies.
    \item (Uniform $\tau$-Reset Strategies): Allocating liquidity uniformly on a range of contiguous buckets centered around the current reference bucket $B_{Z_t}$ and resetting when prices move outside of this range. 
    \item (Context-Independent Allocation Strategies): Setting $A(S,i) = A_i \in \mathbb{R}$ for all $S \in \cS$; i.e.,  the proportional allocations relative to baseline bucket index are always the same at the time of a reset trigger. 
\end{enumerate}

\section{Optimizing Earnings}
\label{sec:optimization}

In this section, we formulate the earnings optimization problem faced by an LP with  {\em belief, $\cP$,} defining a  distribution on contract-market price sequences in a given time horizon. 
In the most general case, belief $\cP$ would depend on the liquidity allocation strategy used by an LP. For example, if the LP provides a large amount of liquidity for a given price interval, this would in turn reduce the slippage of trades at those prices, which may in turn increase the volume of trades facilitated by the contract, and hence change $\cP$. As in~\cite{fan2022differential}, we make the simplifying assumption, reasonable for a small LP, that their  belief $\cP$ is a {\it liquidity-independent distribution}, and independent of the strategic liquidity strategy used by the LP. Going forward, we limit our attention to liquidity-independent beliefs. In particular, we will model non-arbitrage traders who trade to a particular buy or sell contract price whose value is unaffected by this LP's liquidity allocation, along with arbitrage traders whose trades are triggered by considerations of market price vs contract price.

\subsection{Optimal \texorpdfstring{$\tau$}{τ}-reset Strategies}
\label{sec:optimal-tau-reset-strategies}

We've seen that $\tau$-reset strategies are a versatile framework for dynamic liquidity provision. For  a given value of $\tau$, the only choice in defining a $\tau$-reset strategy is  the allocation function $A$, and we let $\Lambda_\tau(A) = (g_\tau,h_\tau,A)$ denote the resulting $\tau$-reset strategy.

In this section, we provide a means of optimizing expected earnings for a given $\tau$.
For this, we assume  that $A \in \cA$, where $\cA$ is a  family of allocation functions. The space of all allocation functions is  large, with an allocation function potentially depending on the entire history of contract-market price sequences and LP actions up to the point when a liquidity reset is triggered. 
  
In defining an LP's optimization problem, we consider LPs with different levels of {\em risk-aversion}, encoded by a {\em utility function},  $u: \mathbb{R} \rightarrow \mathbb{R}$ (we provide example utility functions below), and we assume that an LP wants to select an allocation function to maximize $V^u_{\tau,\cP}(A) = \mathbb{E}_{\vect{P} \sim \cP}  \left[ u(V(\Lambda_\tau(A), \vect{P}) \right]$. With this in hand, we let 
$$
\mathit{OPT}(\tau,\cP,u,\cA) = \max_{A \in \cA }V^u_{\tau,\cP}(A),
$$
and  denote an allocation function in family $\cA$ that achieves optimal earnings by $A^*$. 

In general, convex $u$ and concave $u$ correspond to risk-seeking and risk-averse LPs, respectively, and linear $u$ corresponds to a risk-neutral LPs (where we can adopt $u(x) = x$ without loss of generality). Going forward, we adopt as the utility function that  with constant Arrow-Pratt measures of absolute risk-aversion~\cite{arrow1965aspects, pratt1978risk}.
\begin{definition}[Constant Absolute Risk Aversion Utility]
For a given $a \in \mathbb{R}$, the {\em constant absolute risk aversion utility function},
 $u_a:\mathbb{R} \rightarrow \mathbb{R}$, is 
given by: 
\begin{align}
    u_a(x) &= 
    \begin{cases}
    \left(1-e^{-ax}\right) / a & \mbox{if $a\neq 0$, and} \\
    x & \mbox{otherwise.}
    \end{cases}
\end{align}
For $a < 0$, $a = 0$, and $a > 0$, utility function $u_a$ models a risk-seeking, risk-neutral, and risk-averse agent, respectively. 
\end{definition}

\subsection{Sampling to Approximate \texorpdfstring{$OPT$}{OPT}}

In order to optimize $V^u_{\tau,\cP}(A)$, we approximate the objective by taking a discrete sample of paths from $\cP$. As such, suppose that $\vect{P}_1,\dots,\vect{P}_N \sim \cP$. We define the empirical average earnings of an LP given the sample paths as: 
$$
\hat{V}^u_{\tau}(A \mid \vect{P}_1,\dots,\vect{P}_N) = \frac{1}{N} \sum_{q=1}^N u(V(\Lambda_\tau(A), \vect{P}_q).
$$ 
In expectation, we obtain:
$$
\mathbb{E}_{\vect{P}_1,\dots,\vect{P}_N \sim \cP}  \left[ \hat{V}^u_{\tau}(A \mid \vect{P}_1,\dots,\vect{P}_N) \right] = V^u_{\tau,\cP}(A).
$$
Going forward, we approximate $\mathit{OPT}(\tau,\cP,u,\cA)$ by taking sufficiently many samples from $\cP$ and optimizing $\hat{V}^u_{\tau}$.

\subsection{Computing Optimal \texorpdfstring{$\tau$}{tau}-reset Strategies with Neural Networks}
\label{sec:computing-opt-with-NN}

We compute optimal $\tau$-reset strategies by letting the allocation function, $A$, be parametrized by a feedforward neural network (NN)  with parameters given by $\theta \in \vect{\theta}$. We let $A_\theta$ denote the specific allocation function for a given parameter choice $\theta \in \vect{\theta}$ and we let $\cA_{\vect{\theta}}$ denote the space of all possible parametric settings of the NN. Our objective is to maximize $\hat{V}^u_{\tau}(A_\theta \mid \vect{P}_1,\dots,\vect{P}_N)$ for a given sample of contract/market price paths $\vect{P}_1,\dots,\vect{P}_N \sim \cP$. 

When a reallocation is triggered at the beginning of epoch $j (j = 1, 2, \dots, k)$, the NN takes as input a set of features $C^j$ that contains context information. The set of context features includes the current time step, the current wealth, the current pool price, the current bucket that the pool price lies in, and an exponentially-weighted moving average (the smoothing parameter value is $0.1$) of non-arbitrage trade volume that a hypothetical $1$ unit of liquidity over the entire price range would have achieved given the price trajectory.

We use a fully connected neural network architecture with 5 hidden layers, and the size of each hidden layer is $m = 16$. The size of the input layer is $n=5$ (the number of context features), and the output layer is of size $s = 2\tau + 2$ (the first $2\tau + 1$ dimensions are the proportional capital to be allocated into the corresponding buckets, and there is also a special dimension for not allocating some of the wealth if needed). All the hidden layers are associated with the ReLU activation function. In addition, a soft-max function is added for the final output in order to produce a vector of sum $1$. The architecture we use is visualized in the right image of Figure~\ref{fig:NN-architecture}.

Unpacking the objective, $\hat{V}^u_{\tau}(A_\theta \mid \vect{P}_1,\dots,\vect{P}_N)$ shows a fundamental recurrence in the given allocation function $A_\theta$. This is because an allocation produced by $A_\theta$ is deployed into the pool and affects the value of wealth when the next reallocation is triggered, and wealth is used as part of the input to $A_\theta$ for the new reallocation as visualized in the left diagram of Figure~\ref{fig:NN-architecture}. However, the NN representation of $A$ allows gradients to be pushed through the recurrence with standard back propagation methods used for recurrent neural networks. Given this, we find optimal $\theta \in \vect{\theta}$ via standard gradient descent methods. 

In more detail, for optimization of the NN (ODRA) and the constant allocation (OIRA)\footnote{In Appendix \ref{appendix:convex-optimization} we also provide a natural variant of OIRA for LPs exhibiting risk-aversion via logarithmic utilities as in \cite{cartea2022decentralised}. For this model we provide convex optimization methods to solve for optimal allocations.}, we use stochastic gradient descent based on sampled price trajectories. The number of training steps is 10000 for both optimize methods. The learning rate for the NN is $10^{-3}$ while the learning rate for the constant allocation is $10^{-2}$. In addition, the Adam optimizer is used for both methods.\footnote{The codebase we use to run experiments is open-sourced at \url{https://github.com/Evensgn/uniswap-active-lp}.}

As risk aversion parameter $a$ increases, the relative difference between the utility values of two wealth values $(u_a(x_1) - u_a(x_2)) / u_a(x_1)$ becomes smaller and this could pose a challenge to the optimization of ODRA and OIRA when the improvement of utility value is numerically very small. To resolve this issue, we apply a positive affine transformation to the utility values as $u_a^{*}(x) = (u_a(x) - u_a(1)) / (u_a(1.1) - u_a(1))$ for all $x$ and use the transformed utility values $u_a^{*}(x)$ in the loss function during training of ODRA and OIRA. This helps the optimization process and at the same time does not alter the problem formulation of the optimization as utility functions $u_a$ and $u_a^{*}$ represent the same set of underlying preferences.

\begin{figure}[ht]
    \begin{center}
    \hspace{-4.5cm}
  \begin{minipage}{0.3\textwidth}
    \centering
        \scalebox{0.6}{%
\begin{tikzpicture}
  \node[circle, draw, fill=gray!10] (C) at (-6,0) {$C$};
  \node[circle, draw] (A) at (-6,2) {$A_\theta$};
  \node[circle, draw, fill=gray!10, minimum size=0.85cm] (x) at (-6,4) {$x$};
  \node[circle, draw, fill=gray!10] (W) at (-6,6) {$W$};

  \node[circle, draw, fill=gray!10] (W0) at (-2,6) {$W^0$};
  
  \node[circle, draw, fill=gray!10] (C0) at (0,0) {$C^0$};
  \node[circle, draw] (A0) at (0,2) {$A_\theta$};
  \node[circle, draw, fill=gray!10] (x0) at (0,4) {$x^0$};
  \node[circle, draw, fill=gray!10] (W1) at (0,6) {$W^1$};

  \node[circle, draw, fill=gray!10] (C1) at (2,0) {$C^1$};
  \node[circle, draw] (A1) at (2,2) {$A_\theta$};
  \node[circle, draw, fill=gray!10] (x1) at (2,4) {$x^1$};
  \node[circle, draw, fill=gray!10] (W2) at (2,6) {$W^2$};

  \node[circle, draw, fill=gray!10] (C2) at (4,0) {$C^2$};
  \node[circle, draw] (A2) at (4,2) {$A_\theta$};
  \node[circle, draw, fill=gray!10] (x2) at (4,4) {$x^2$};
  \node[circle, draw, fill=gray!10] (W3) at (4,6) {$W^3$};

  \node[circle, draw, fill=gray!10] (Ck) at (8,0) {$C^k$};
  \node[circle, draw] (Ak) at (8,2) {$A_\theta$};
  \node[circle, draw, fill=gray!10] (xk) at (8,4) {$x^k$};
  \node[circle, draw, fill=green!10] (Wk1) at (8,6) {$W^{k+1}$};

  \draw[very thick,-latex] (C) -- (A);
  \draw[very thick,-latex] (A) -- (x);
  \draw[very thick,-latex] (x) -- (W);
  
  \draw[very thick,-latex] (C0) -- (A0);
  \draw[very thick,-latex] (A0) -- (x0);
  \draw[very thick,-latex] (x0) -- (W1);

  \draw[very thick,-latex] (C1) -- (A1);
  \draw[very thick,-latex] (A1) -- (x1);
  \draw[very thick,-latex] (x1) -- (W2);

  \draw[very thick,-latex] (C2) -- (A2);
  \draw[very thick,-latex] (A2) -- (x2);
  \draw[very thick,-latex] (x2) -- (W3);

  \draw[very thick,-latex] (Ck) -- (Ak);
  \draw[very thick,-latex] (Ak) -- (xk);
  \draw[very thick,-latex] (xk) -- (Wk1);

  \draw[very thick,-latex,rounded corners] (W0) -- (-1,6) -- (-1,0) -- (C0);
  \draw[very thick,-latex,rounded corners] (W1) -- (1,6) -- (1,0) -- (C1);
  \draw[very thick,-latex,rounded corners] (W2) -- (3,6) -- (3,0) -- (C2);
  \draw[very thick,dotted,-latex] (W3) -- (5.5,6);
  \draw[very thick,dotted,-latex] (6.5,0) -- (Ck);

  \draw[very thick,-latex,rounded corners] (W) -- (-5,6) -- (-5,0) -- (C);

  \node[scale=2] (eq) at (-3.5,3) {$=$};

  \node[scale=2] (dots) at (6,3) {$\dots$};
\end{tikzpicture}
        }
    \vspace{0.05cm}
    \label{fig:recurrent-NN}
  \end{minipage}%
  \end{center}
  \hspace{4mm}
  \begin{minipage}{0.35\textwidth}
    \centering
        \scalebox{0.65}{%
        \begin{tikzpicture}[x=3.2cm,y=1.4cm]
          \message{^^JNeural network, shifted}
          \readlist\Nnod{4,5,5,5,5,5,3} 
          \readlist\Nstr{n,m,m,m,m,m,s} 
          \readlist\Cstr{\strut c,a^{(\prev)},a^{(\prev)},a^{(\prev)},a^{(\prev)},a^{(\prev)},x} 
          \def\yshift{0.5} 
          
          \message{^^J  Layer}
          \foreachitem \N \in \Nnod{ 
            \def\lay{\Ncnt} 
            \pgfmathsetmacro\prev{int(\Ncnt-1)} 
            \message{\lay,}
            \foreach \i [evaluate={\c=int(\i==\N); \y=\N/2-\i-\c*\yshift;
                         \index=(\i<\N?int(\i):"\Nstr[\lay]");
                         \x=\lay; \n=\nstyle;}] in {1,...,\N}{ 
              \node[node \n] (N\lay-\i) at (\x,\y) {$\Cstr[\lay]_{\index}$};
              
              \ifnum\lay>1 
                \foreach \j in {1,...,\Nnod[\prev]}{ 
                  \draw[connect,white,line width=1.2] (N\prev-\j) -- (N\lay-\i);
                  \draw[connect] (N\prev-\j) -- (N\lay-\i);
                }
              \fi 
              
            }
            \path (N\lay-\N) --++ (0,1+\yshift) node[midway,scale=1.5] {$\vdots$};
          }
          
          \node[above=5,align=center,mygreen!60!black] at (1,-3.5) {input\\[-0.2em]layer};
          \node[above=2,align=center,myblue!60!black] at (4, 0) {hidden layers};
          \node[above=10,align=center,myred!60!black] at (7,-9) {output\\[-0.2em]layer};  
        \end{tikzpicture}
        }
    \label{fig:feedforward-NN}
  \end{minipage}
\caption{The top image provides a visualization of the recurrence in $A_\theta$ for the overall objective $\hat{V}^u_{\tau}(A_\theta \mid \vect{P}_1,\dots,\vect{P}_N)$ which we exploit to compute gradients in a similar fashion to recurrent neural networks. In this image, $C$ denotes the context that is fed to the neural network $A_\theta$ as features. A relevant feature at each epoch is the wealth that the LP has accumulated at the beginning of the epoch $W^i$, which is exemplified via an arrow in the figure. The overall objective is the given utility function applied to the wealth at the end of the final epoch $W^{k+1}$. The bottom image provides a visualization of the neural network architecture we use for $A_\theta$. There is a soft-max function applied to the output layer. The recurrent structure of the objective's dependence with respect to the NN parameters, $\theta  \in \vect{\theta}$ allow us to use techniques from recurrent neural networks to compute the gradient of the objective $u(W^{k+1})$ with respect to $\theta$. 
\label{fig:NN-architecture}}
\end{figure}

\subsection{Liquidity Provision Strategies}

Below we outline the main strategies we compare in different regimes:  
\begin{enumerate}
    \item Optimal static allocation (OSA): This strategy computes $\vect{x}$ that optimizes the value of $u_a(V(\vect{x} \mid \vect{P}_1,\dots,\vect{P}_N))$ from Section~\ref{sec:static-LP-earnings}. This is the only strategy that does not explicitly use liquidity reallocations (though it can be seen as a $\tau$-reset strategy with $\tau > T$), and it coincides with the work of \cite{fan2022differential}.  
    \item Uniform liquidity $\tau$-reset allocation (ULRA): For a fixed $\tau$, this strategy mints an equal $\mu$ units of liquidity for each of the $2\tau + 1$ contiguous buckets considered in a reallocation. $\tau$ is chosen to be as large as possible so the LP makes use of the entire wealth at their disposal at a reset to reallocate liquidity. 
    \item Uniform proportional $\tau$-reset allocation (UPRA): For a fixed $\tau$, this allocates wealth in equal proportions to each of the $2\tau + 1$ buckets after a reset (in general this does not result in a uniform liquidity allocation as the cost per unit of liquidity in each bucket may be different). 
    \item Optimal context-independent $\tau$-reset allocation (OIRA): For a fixed $\tau$, this computes the optimal single allocation vector to be used at every reset. In other words, the LP computes an optimal $(A_{-\tau},\dots,A_{\tau})$, to be used to allocate liquidity around the reference bucket at each reallocation.
    \item Optimal context-dependent $\tau$-reset allocation (ODRA): For fixed $\tau$, this is solved with the Neural Network formulation of Section~\ref{sec:computing-opt-with-NN}.
\end{enumerate}

\section{Experimental Setup: Contract-Market Prices}
\label{sec:experimental-setup}

In this section, we describe a family of empirically-informed 
contact-market price sequences against which we will optimize $\tau$-reset strategies
and we use historical data to inform this stochastic price model. 

\subsection{Modeling Contract-Market Prices}
\label{sec:Modeling-Contract-Market-Prices}

For this, we use a similar approach to \cite{fan2022differential}, which is in turn inspired by \cite{capponi2021adoption}, to provide a family of liquidity-independent contract-market price distributions. This makes use of an external stochastic process to define a sequence of market prices, together with 
non-arbitrage trades that affect the contract price and arbitrage trades that act to
bring contract prices closer to market prices.

We assume that contract-market prices are generated over each of $R > 0$ {\it rounds}. At the beginning of each round, market prices change randomly according to a stochastic process $\cP_M$. During the $r$-th round, after the contract-market price update, there are some number, $k_r \geq 0$, of  non-arbitrage trades which impact contract price, $P_c$, only.  Each non-arbitrage trade is either a purchase or a sale, this determined uniformly at random with probability $1/2$. The effect of such a trade is that the contract price changes to $(1+\lambda_r)P_c$ or $(1+\lambda_r)^{-1}P_c$ respectively, where $\lambda_r > 0$, depending on whether a purchase or sale occured. Crucially, trades are price-based in our model rather than volume based, which in turn ensures that contract-market prices evolve independent of liquidity provided by an LP. It is precisely this exogenous uncertainty to LP actions that will allows us to compute optimal $\tau$-reset strategies via the methods of Section~\ref{sec:optimization}, as this allows us to sample price paths first and then optimize LP allocation functions.

We also model  arbitrage trades whose role is to bring contract prices close to market prices. 
For this, we follow~\cite{fan2022differential}, and with a  Uniswap contract fee rate, $\gamma \in (0,1)$, we let $I_\gamma(P_m) = [(1-\gamma)P_m, (1-\gamma)^{-1}P_m]$ be the {\em no-arbitrage interval} around the market price $P_m$. If the contract price exits this no-arbitrage interval, we assume a arbitrage trade brings the contract price to the closest price in the interval. That is, if $P_c <  (1-\gamma)P_m$ we assume that arbitrage trade moves the contract price to $(1-\gamma)P_m$, and  if $P_c > (1-\gamma)^{-1}P_m$ we assume that arbitrage trade moves the contract price to $(1-\gamma)^{-1}P_m$.
\begin{definition}[Round-Based Liquidity-Independent Price Distribution]
\label{def:round-based-liq-ind-dist}
We say that $\cP$ is a  {\em round-based liquidity-independent price distribution} when it is a  distribution that  is parameterized by: 
\begin{itemize}
\item $R > 0$: the number of rounds,
\item $\gamma \in (0,1)$: the fee rate of the Uniswap contract,
\item $\cP_M$: the stochastic process governing market price updates at the beginning of each round,
\item $\vect{k} = (k_r)_{r=1}^R$ with $k_r > 0$: the number of non-arbitrage trades in
 each round $r \in \{1,\dots,R\}$, and
\item $\vect{\lambda} = (\lambda_r)_{r=1}^R$ with $\lambda_r > 0$: the multiplicative impact of a non-arbitrage trade on contract price for each round $r \in \{1,\dots,R\}$.
\end{itemize}

When we wish to specify the resulting round-based price distribution, we write this as $\cP(R,\gamma,\cP_M,\vect{k},\vect{\lambda})$.
\end{definition}

We model $\cP_M$ as a geometric Brownian motion with parameters estimated from historical price data between token pairs. We also explore multiple regimes of time-varying non-arbitrage trade by varying $\vect{\lambda}$ (the framework  is flexible enough to permit arbitrary values of $\lambda_r$ for each round).

\subsection{Market Prices as a Geometric Brownian Motion}

We model the stochastic nature of market prices, $\cP_M$, as a Geometric Brownian Motion (GBM).
If the time series is given by $X_1,\dots,X_T$, then the successive multiplicative increments of the time series are i.i.d lognormally distributed. If we let $Z_i = \log \left( \frac{X_{i}}{X_{i-1}} \right)$, then $Z_2,\dots Z_T \sim \text{iid } \cN(\mu,\sigma^2)$ with {\em drift}, $\mu$, and {\em diffusion}, $\sigma$. 
We estimate these parameters on  per-minute time series data for ETH/BTC prices (the {\em low volatility regime}) and ETH/USDT (the {\em high volatility regime}) from March 2022 through February 2023. For each time series, we estimate the drift and diffusion via standard MLE methods.
The following are the MLE estimates we obtain for $\mu$ and $\sigma^2$ for each of the two volatility regimes: 
\begin{center}
\begin{tabular}{ |c|c|c| } 
 \hline
  & ETH/BTC & ETH/USDT \\
 \hline 
 $\hat{\mu}$ & $4.835 \times 10^{-8}$ & $-1.140 \times 10^{-6}$ \\
 \hline 
 $\hat{\sigma}^2$ & $1.946 \times 10^{-7}$ & $8.329 \times 10^{-7}$ \\ 
 \hline
\end{tabular}
\end{center}

\begin{figure}[ht]
  \centering
  \begin{subfigure}[b]{0.4\textwidth}
    \centering
    \includegraphics[width=\textwidth]{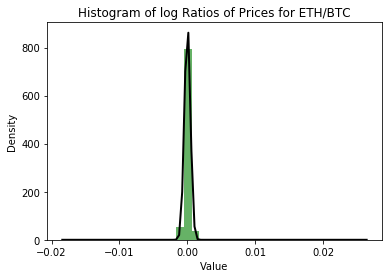}
    \label{fig:eth-btc-log-ratios}
  \end{subfigure}
  \hspace{0.1\textwidth}
  \begin{subfigure}[b]{0.4\textwidth}
    \centering
    \includegraphics[width=\textwidth]{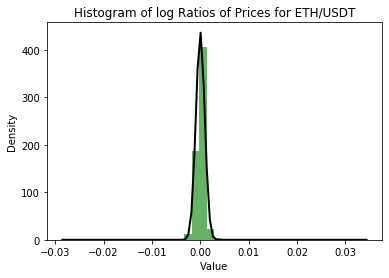}
    \label{fig:eth-usdt-log-ratios}
  \end{subfigure}
  \caption{Log ratios of consecutive prices (per-minute) between different asset pairs. Left: ETH/BTC, representing low-volatility contract-market price sequences as these prices are highly correlated. Parameter estimates give rise to values $\hat{\mu} \approx 4.84 \times 10^{-8}$ and $\sigma^2 \approx 1.95 \times 10^{-7}$.  Right: ETH/USDT, representing high-volatility contract-market price sequences as these prices are less correlated. Parameter estimates give rise to values $\hat{\mu} \approx -1.14 \times 10^{-6}$ and $\sigma^2 \approx 8.33 \times 10^{-7}$
  \label{fig:empirical-log-ratios}}
\end{figure}

\subsection{Contract Price Updates}

Whereas arbitrage trades are specified by the fee rate of the contract, non-arbitrage trades are parametrized by $\vect{k} = (k_r)_{r=1}^R$ and $\vect{\lambda} = (\lambda_r)_{r = 1}^R$, which specify the number and multiplicative magnitude of price change updates arising from non-arbitrage trades in a given round. In our experiments, we fix $k_r = 10$ for each round and introduce time-varying non-arbitrage price flow by  explicitly modulating $\vect{\lambda}$ before sampling from $\cP$. In particular we explore $\vect{\lambda}$ such that $\lambda_r = \bar{\lambda} +  \alpha \cdot tanh(10 (t / T - 0.5))$, where $\bar{\lambda} > 0$ is the average $\lambda_r$ value over the time horizon and $\alpha > 0$ is the variation exhibited in $\lambda_r$ around the mean. 

\section{Experimental Results}
\label{sec:results}

In this section, we explore the increase in earnings that LPs can gain through the use of  dynamic allocation strategies, studying the performance of various liquidity provision strategies in a multitude of economic environments modulated by contract/market price volatility, LP risk-aversion, and reallocation costs. Most importantly, we find many settings in which optimal $\tau$-reset strategies  outperform simpler liquidity provision strategies. In all the experiments that follow, we assume a default setting of $(W,\gamma,R,k_r,\bar{\lambda},\alpha, \eta) = (1,0.003,1000,10,0.00005,0.00005, 0.01)$. When deviating from the default setting we clarify which parameters are changed. In addition, we assume that the buckets of the v3 contract $\vect{\mu} = \{B_{-m},..,B_n\}$ are given by $B_i = [a_i,b_i] = [\phi^i,\phi^{i+1}]$ for $\phi = 1.0001^{10}$.

\subsection{The Impact of Price Volatility}

\begin{figure}
  \centering
  \includegraphics[width=\textwidth]{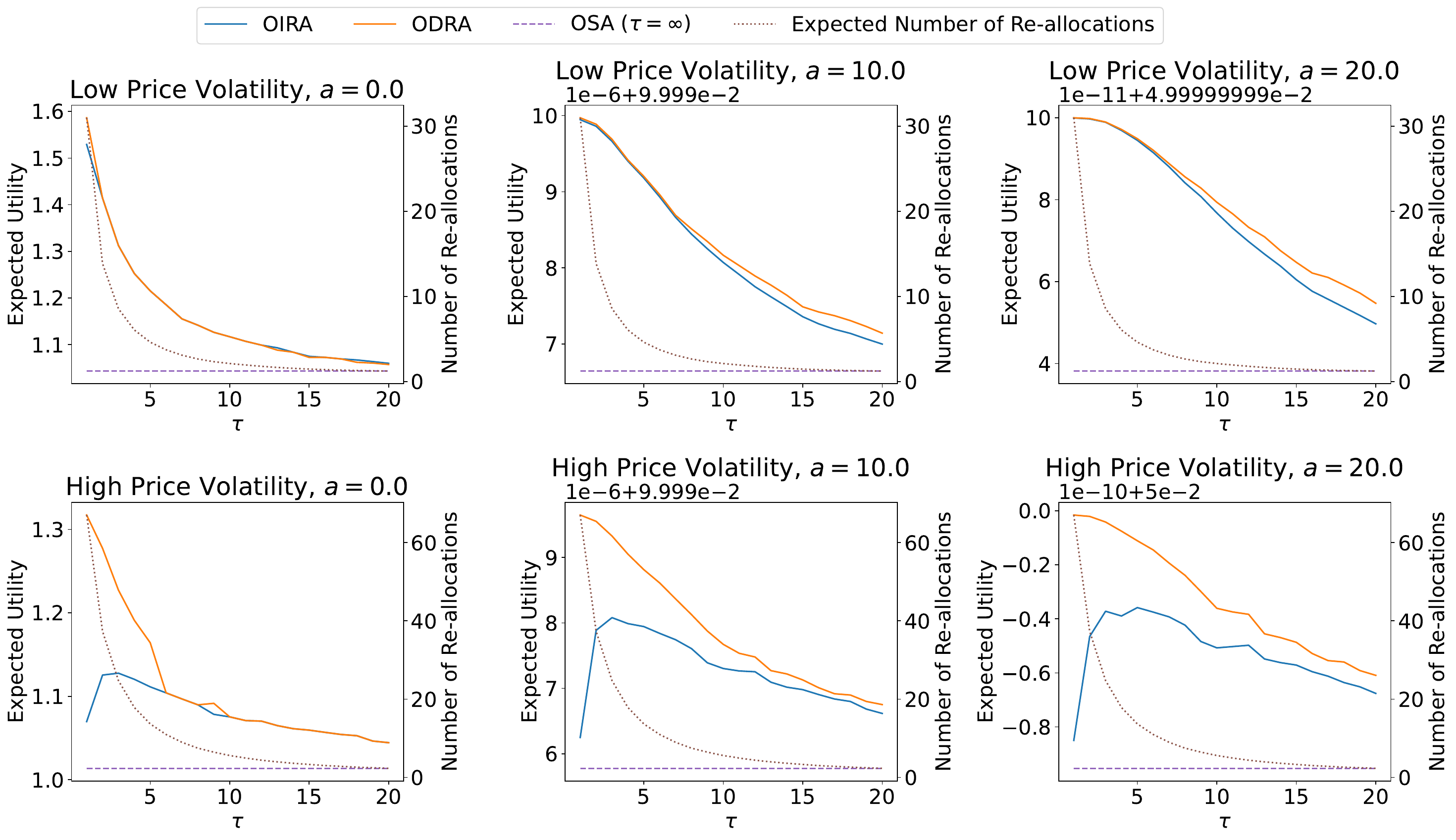}
  \caption{The performance of OIRA, ODRA and OSA strategies as we modulate both risk-aversion and $\cP_M$. For each strategy, we plot the expected utility it achieves as a function of $\tau$, and we also plot the expected number of reallocations that occur as a function of $\tau$. The top row corresponds to a low volatility $\cP_M$, empirically informed from ETH/BTC prices, and the bottom row corresponds to high volatility $\cP_M$, empirically informed from ETH/USDC prices. The columns correspond to risk-aversion values $a = 0,10$, and $20$, respectively from left to right. When scientific notation is used for the y-axis values in certain subplots, it is denoted by a number above the respective y-axis.
  \label{fig:low_vs_high_volpm_utils_subset}}
\end{figure}

\begin{figure}
  \centering
  \includegraphics[width=\textwidth]{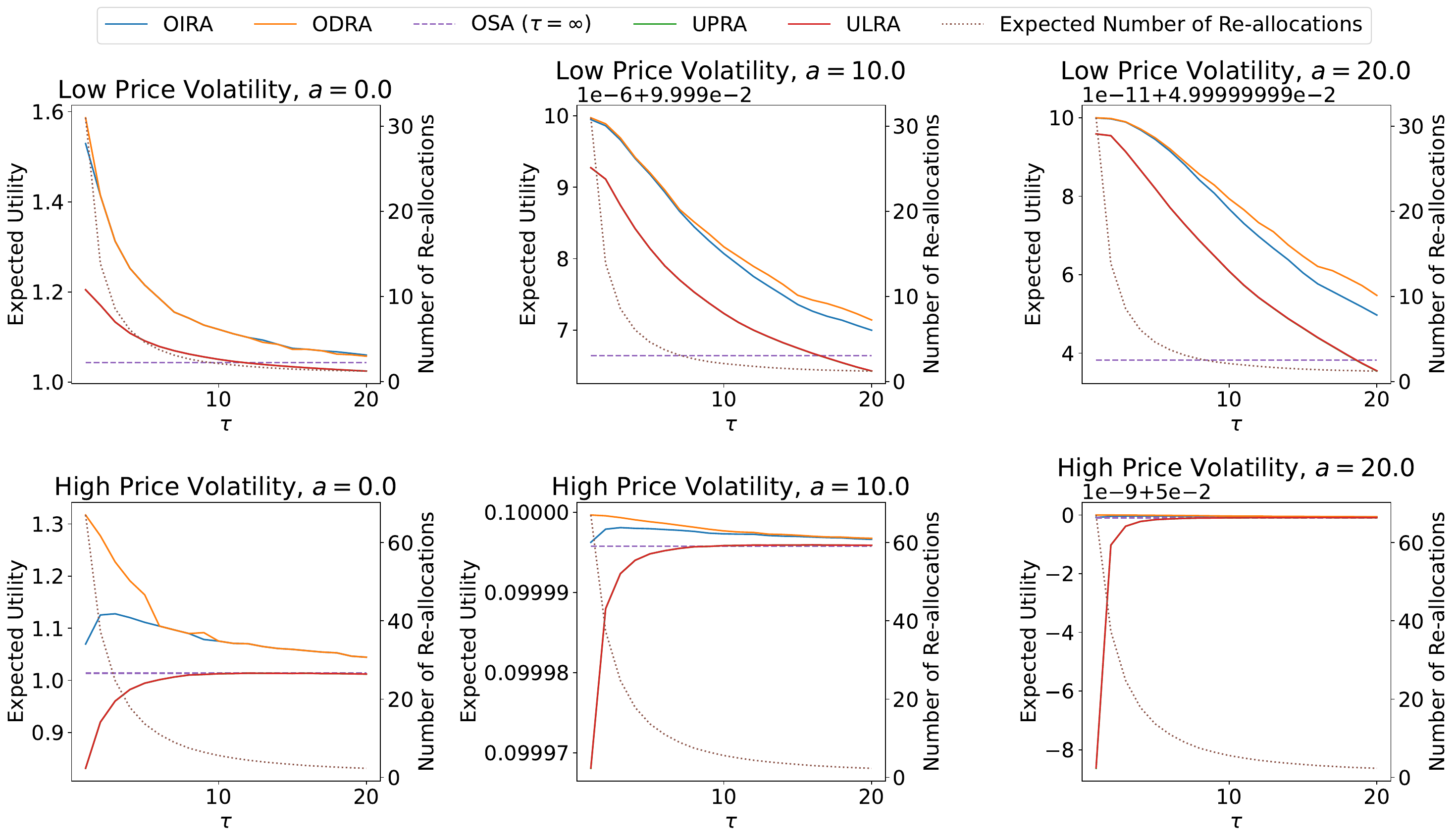}
  \caption{The performance of all strategies as we modulate both risk-aversion and $\cP_M$. For each strategy we plot the expected utility it achieves as a function of $\tau$, and we also plot the expected number of reallocations that occur as a function of $\tau$. The top row corresponds to a low volatility $\cP_M$, empirically informed from ETH/BTC prices, and the bottom row corresponds to high volatility $\cP_M$, empirically informed from ETH/USDC prices. The columns correspond to risk-aversion values $a = 0,10$, and $20$, respectively from left to right. When scientific notation is used for the y-axis values in certain subplots, it is denoted by a number above the respective y-axis.
  \label{fig:low_vs_high_volpm_utils_all}}
\end{figure}

In Figures~\ref{fig:low_vs_high_volpm_utils_subset} and~\ref{fig:low_vs_high_volpm_utils_all} we plot the performance of all LP strategies as we modulate $\cP_M$ from low to high volatility as well as the risk-aversion of the LP. Figure~\ref{fig:low_vs_high_volpm_utils_subset} focuses on only comparing OIRA, ODRA and OSA to tease out the relative performance of ODRA vs. OIRA. Figure~\ref{fig:low_vs_high_volpm_utils_all} incorporates UPRA and ULRA, from where we can see that their performance is almost identical.
The NN-based ODRA outperforms all strategies, especially OSA which does not make use of reallocations. As risk-aversion increases, we see that the distinction between ODRA and OIRA becomes more clear in the plots, however, this does not imply a greater magnitude of performance due to the fact that different risk-aversion values give rise to different scales. In addition, we see that OIRA generally exhibits optimal performance with $\tau > 1$ whereas all other $\tau$-reset strategies in this setting perform better with $\tau = 1$. 

\begin{figure}
  \centering
  \includegraphics[width=\textwidth]{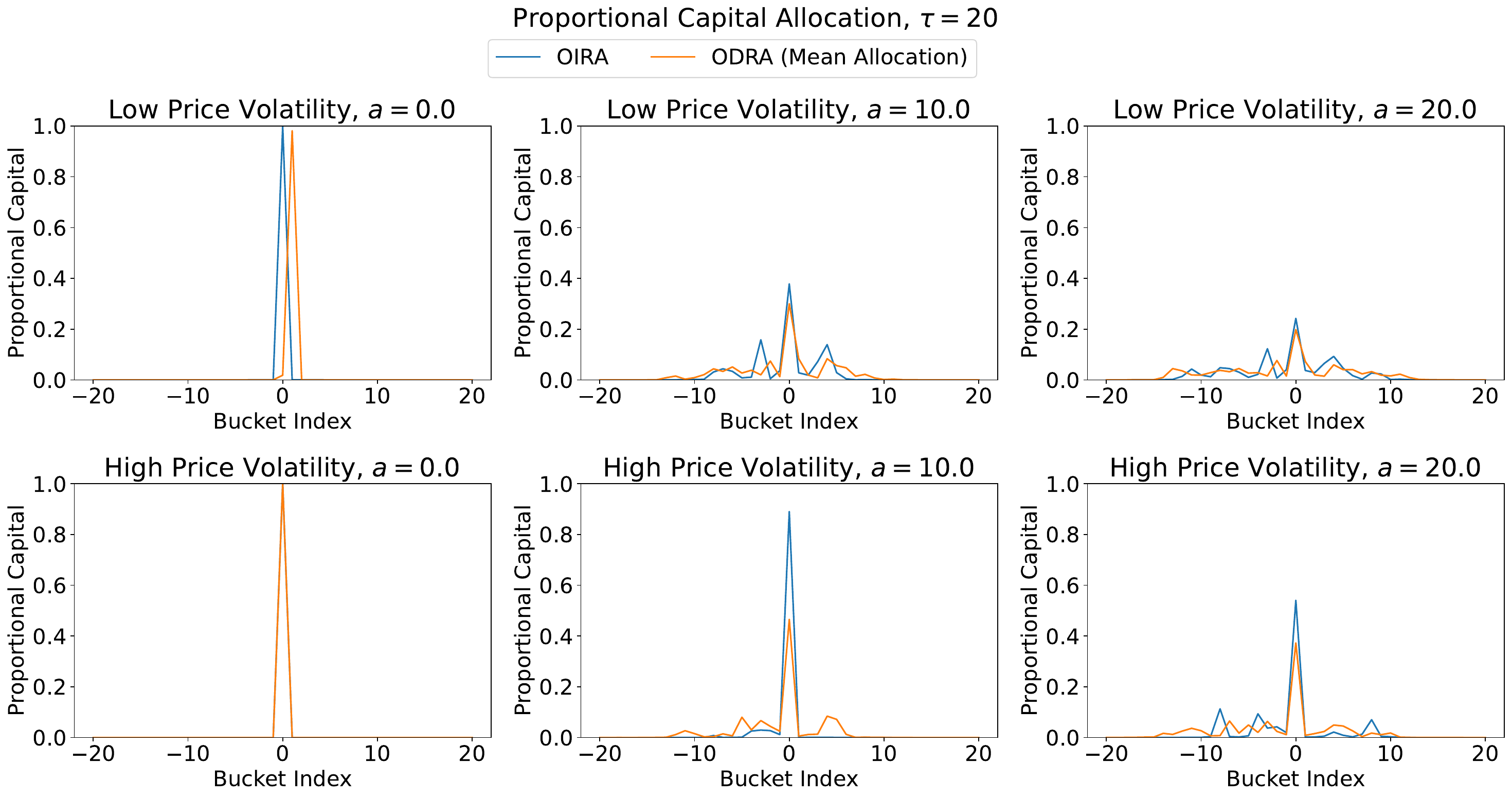}
  \caption{OIRA allocation and ODRA average allocations for $\tau = 20$ as we modulate both risk-aversion and $\cP_M$. The top row corresponds to a low volatility $\cP_M$, empirically informed from ETH/BTC prices, and the bottom row corresponds to high volatility $\cP_M$, empirically informed from ETH/USDC prices. The columns correspond to risk-aversion values $a = 0,10$, and $20$, respectively from left to right. 
  \label{fig:low_vs_high_volpm_alloc}}
\end{figure}

In terms of the impact of $\cP_M$, we see that for lower $\tau$ values, higher $\cP_M$ leads to a larger separation between ODRA and OIRA in performance. Moreover, Figure~\ref{fig:low_vs_high_volpm_alloc} plots allocation profiles for ODRA and OIRA as we modulate risk-aversion and $\cP_M$. As expected, with higher risk-aversion we see a larger spread in allocations, as LPs may seek to decrease the variance in their earnings with wider positions. On the other hand, as $\cP_M$ increases in volatility, we see that LP positions for both ODRA and OIRA become more narrow. This is likely due to the fact that the expected number of reallocations is higher in the high volatility setting than in the low volatility setting for the same $\tau$. For a lower frequency of reallocations, the allocated liquidity is used for longer time periods, hence an LP may wish to spread liquidity over various buckets. 


\subsection{Varying Non-arbitrage Flow}

\begin{figure}
  \centering
  \includegraphics[width=\textwidth]{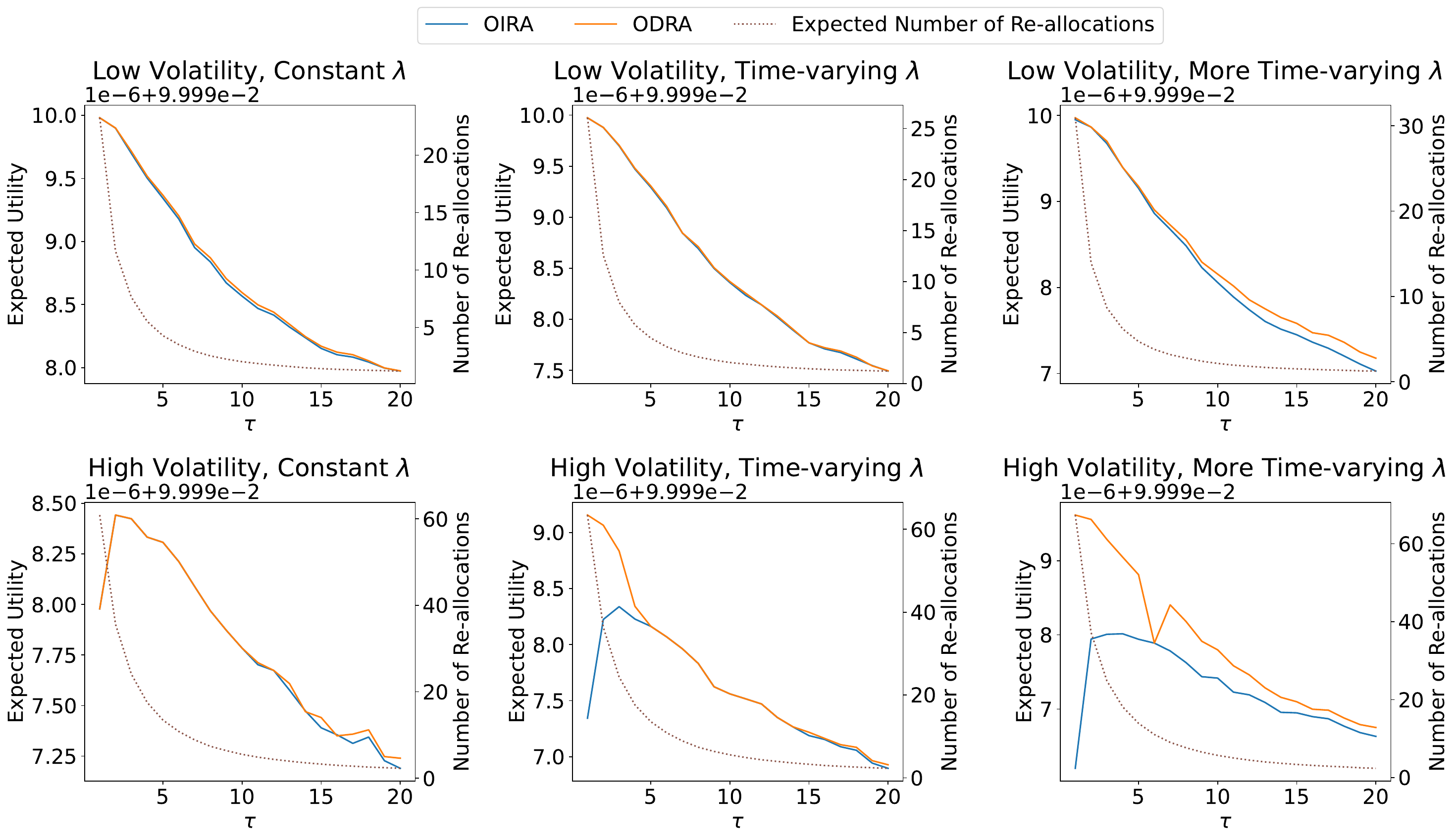}

  \caption{The performance of OIRA and ODRA strategies as we modulate $\cP_M$ and $\vect{\lambda}$. For all plots, we use $a = 10$ for risk aversion and $\bar{\lambda} = 0.00005$ and we modulate the $\alpha$ in $\{0.0, 0.00003, 0.00005\}$ in columns from left to right. The top row plots low volatility $\cP_M$ and the bottom row plots high volatility $\cP_M$. When scientific notation is used for the y-axis values in certain subplots, it is denoted by a number above the respective y-axis.
  \label{fig:adjust-tanh-amplitude-utility}}
\end{figure}

In Figure~\ref{fig:adjust-tanh-amplitude-utility} we modulate the volatility of $\cP_M$ and the magnitude of non-arbitrage flow in $\vect{\lambda}$ by modulating $\alpha$, the amplitude of change in $\vect{\lambda}$ while keeping mean $\vect{\lambda}$ the same. The most salient observation is that as $\vect{\lambda}$ becomes more time-varying, the NN-based approach of ODRA increases its performance relative to OIRA. This is to be expected due to the fact that ODRA can incorporate temporal context in deciding an allocation after a reset, and the non-arbitrage flow inherently has the temporal context of increased importance as $\alpha$ increases. 

\begin{figure}
  \centering
  \includegraphics[width=\textwidth]{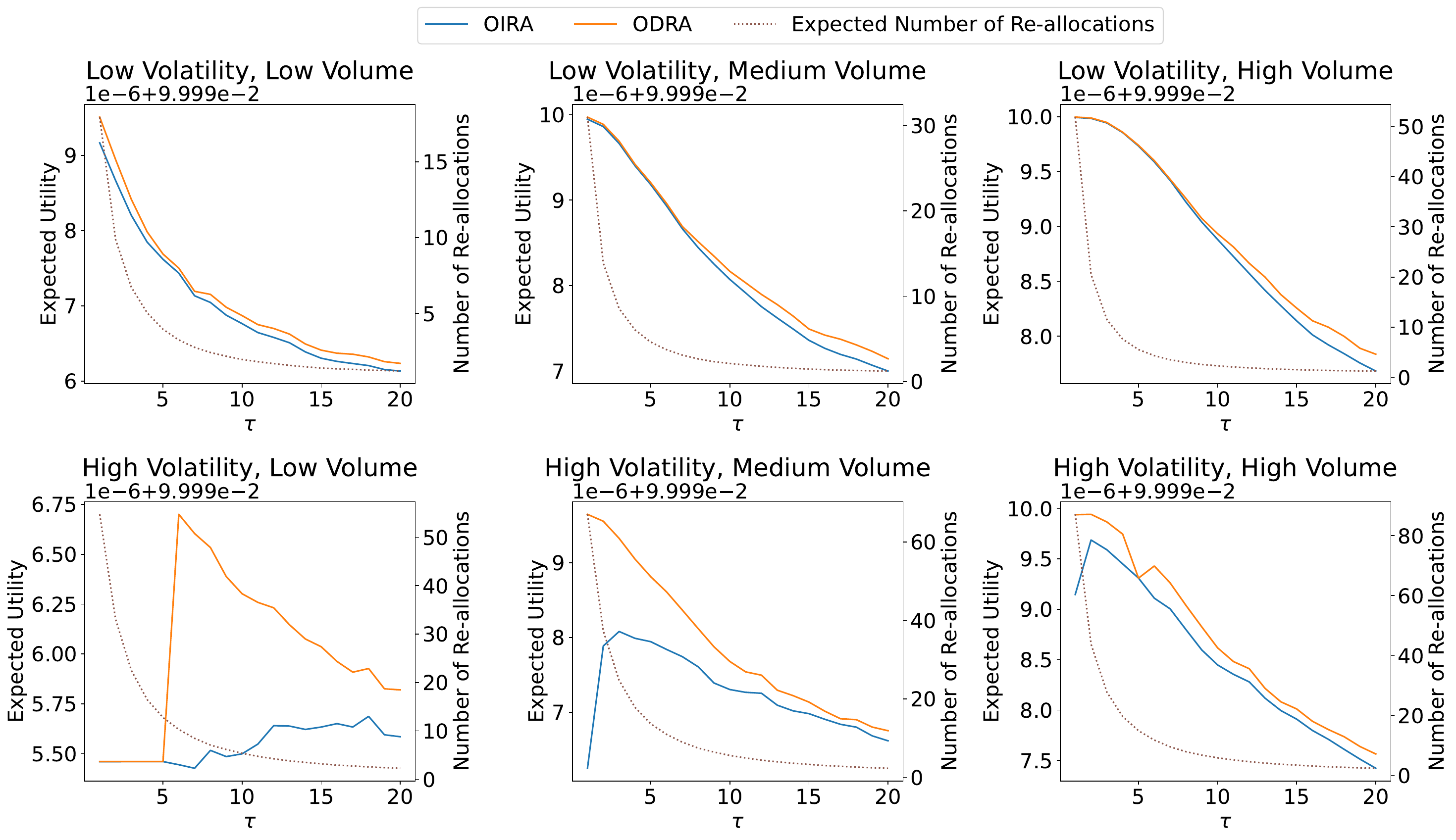}

  \caption{The performance of OIRA and ODRA strategies as we modulate $\cP_M$ and $\vect{\lambda}$. For all plots, we use $a = 10$ for risk aversion and we modulate the $(\bar{\lambda},\alpha) \in \{ (0.000025, 0.000025), (0.00005, 0.00005), (0.000075, 0.000075)\}$  in columns from left to right. The top row plots low volatility $\cP_M$ and the bottom row plots high volatility $\cP_M$. When scientific notation is used for the y-axis values in certain subplots, it is denoted by a number above the respective y-axis.
  \label{fig:adjust-mean-lambda-and-amp-utility}}
\end{figure}

\begin{figure}
  \centering
  \includegraphics[width=\textwidth]{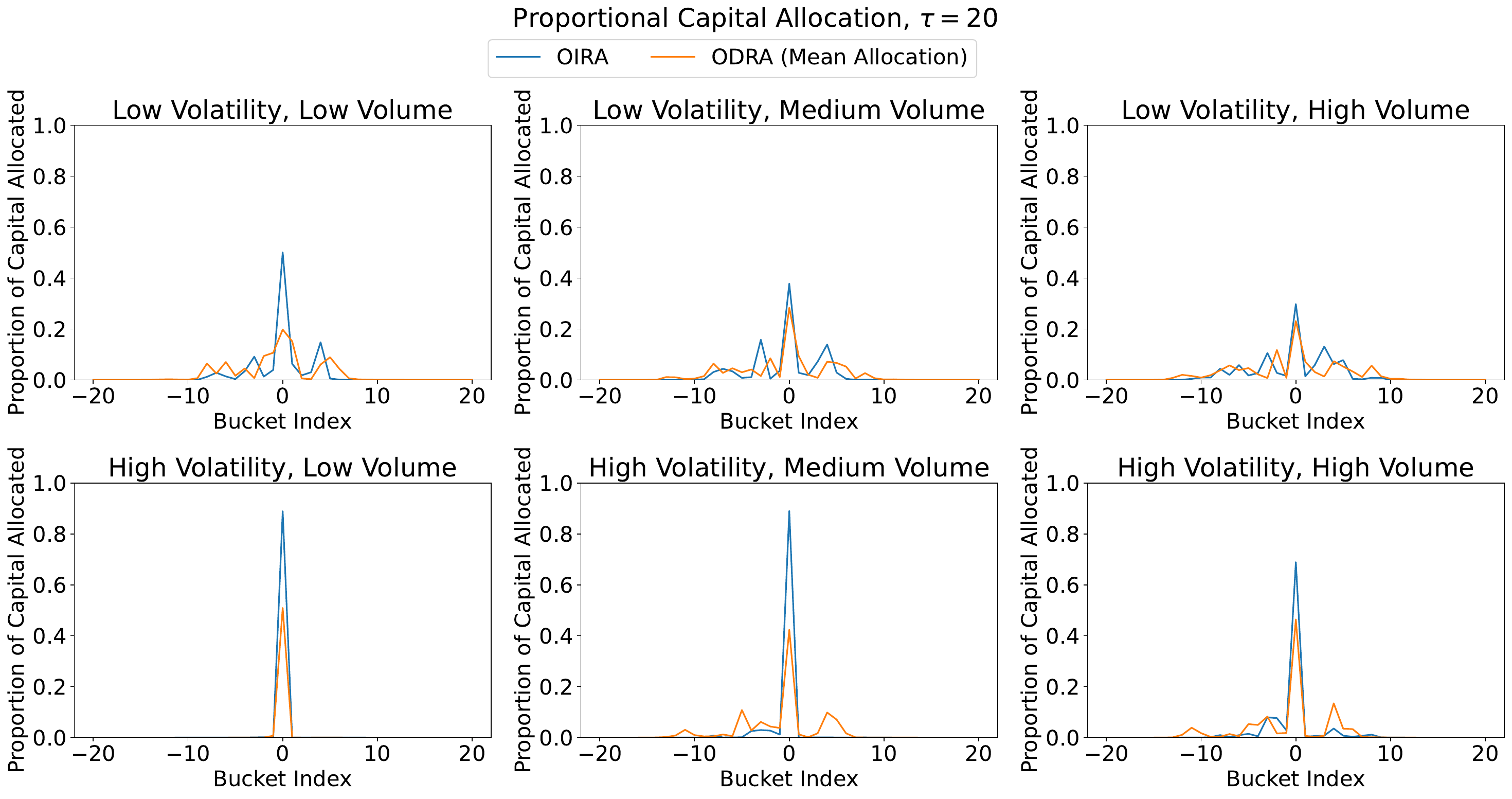}

  \caption{OIRA allocation and ODRA average allocation as we modulate $\cP_M$ and $\vect{\lambda}$. For all plots, we use $a = 10$ for risk aversion and we modulate the $(\bar{\lambda},\alpha) \in \{ (0.000025, 0.000025), (0.00005, 0.00005), (0.000075, 0.000075)\}$  in columns from left to right. The top row plots low volatility $\cP_M$ and the bottom row plots high volatility $\cP_M$. }
  \label{fig:adjust-mean-lambda-and-amp-allocation}
\end{figure}

In Figure~\ref{fig:adjust-mean-lambda-and-amp-utility} and~\ref{fig:adjust-mean-lambda-and-amp-allocation} we also modulate $\vect{\lambda}$ albeit by jointly modulating amplitude($\alpha)$ and mean of $\lambda_r$ values, $\bar{\lambda}$. Once more we see that the NN-based ODRA strategy outperforms all strategies, and we see that the optimal $\tau$ values for ODRA drastically differ in the high volatility $\cP_M$ over those of Figure~\ref{fig:adjust-tanh-amplitude-utility}. Moreover in Figure~\ref{fig:adjust-mean-lambda-and-amp-allocation} we see that both increased non-arbitrage flow and $\cP_M$ volatility contribute to more spread allocations. LPs make profits from non-arbitrage trades, hence they stand to obtain more fees with wider positions for larger flows of non-arbitrage trade. 


\subsection{The Impact of Risk-aversion}

\begin{figure}
  \centering
  \includegraphics[width=\textwidth]{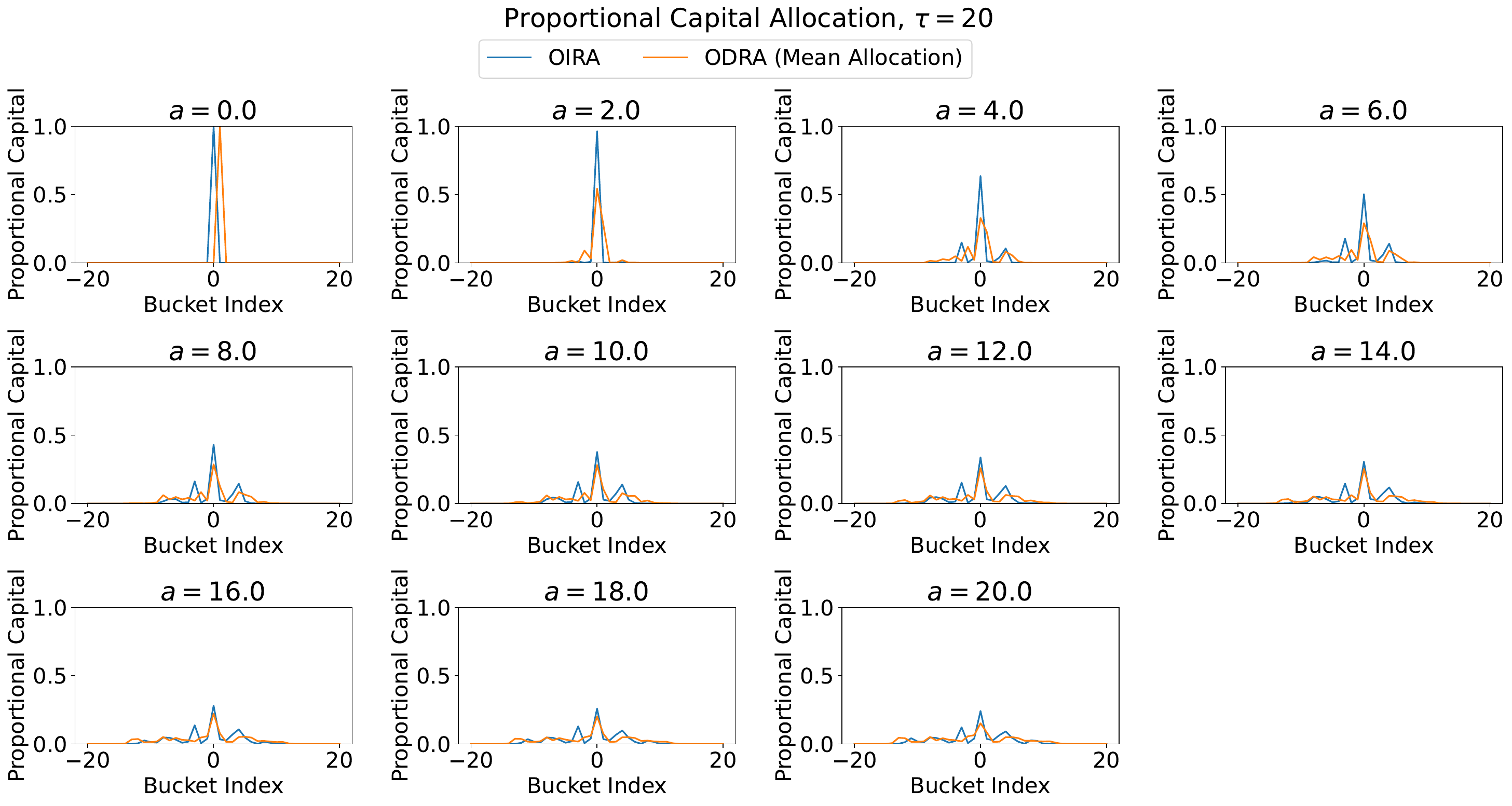}

  \caption{OIRA allocation and ODRA average allocation for $\tau = 20$ for low volatility $\cP_M$ as we modulate risk-aversion from $a = 0$ to $a = 20$. }
  \label{fig:adjust-risk-a-cost-allocation}
\end{figure}

As mentioned in the previous sections, risk-aversion mostly impacts the allocations used in ODRA and OIRA LP strategies. In Figure~\ref{fig:adjust-risk-a-cost-allocation} we make fine-grained modulations of risk-aversion and see that indeed LPs spread their liquidity more as they become more risk-averse. A larger spread of liquidity allocation typically implies lower expected earnings for an LP as they have less proportional liquidity at prices that are traded at, but at the same time, there is less risk of missing out on fees due to prices escaping their position or suffering impermanent loss due to price deviating from the initial price. 

\subsection{The Impact of Reallocation Costs}

\begin{figure}
  \centering
  \includegraphics[width=\textwidth]{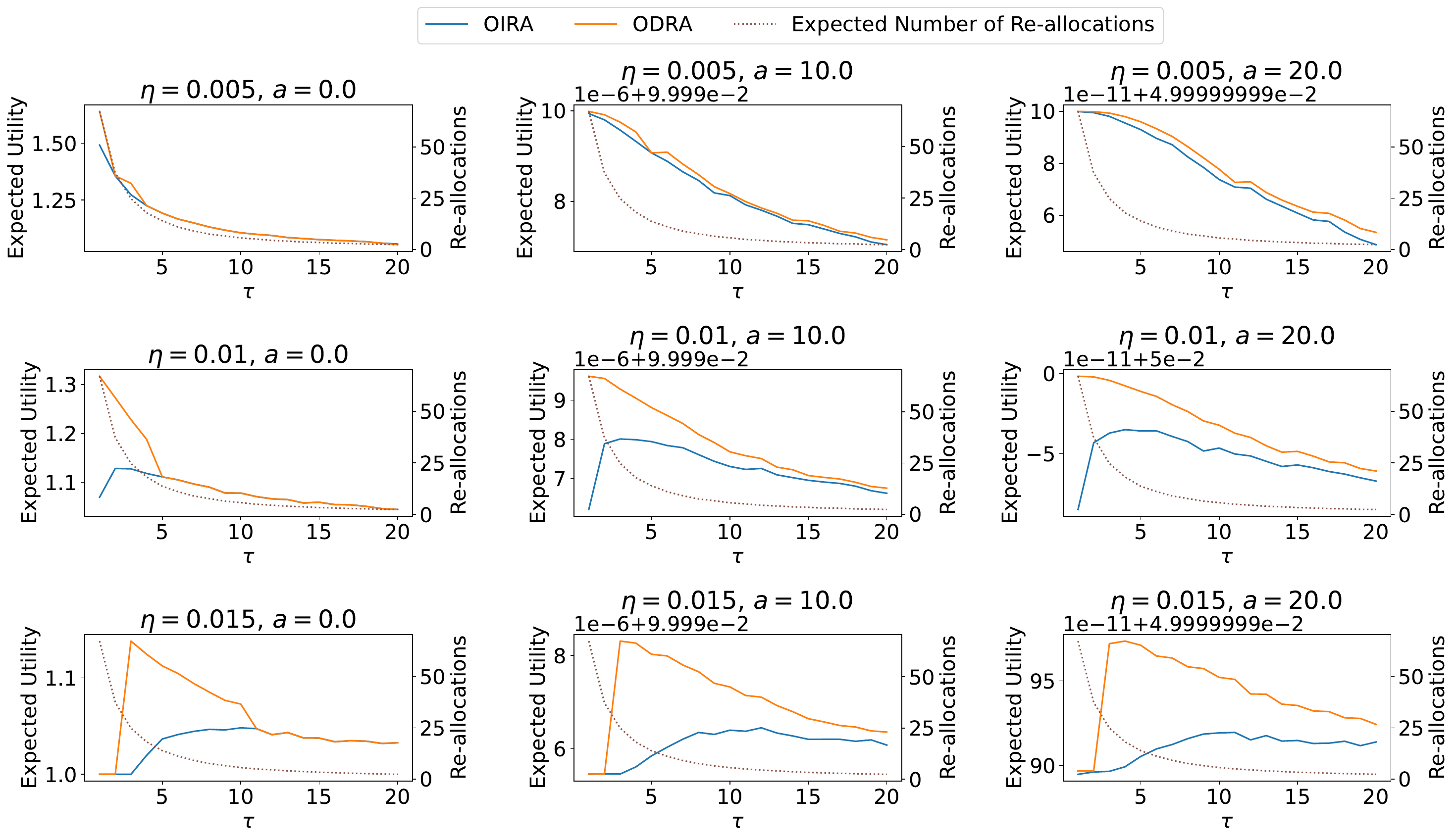}

  \caption{The performance of OIRA and ODRA strategies as we modulate $\eta$ and risk-aversion. For all plots, we use high volatility $\cP_M$. We modulate $a$ in $\{0, 10, 20\}$ in columns from left to right and $\eta \in \{0.005, 0.01, 0.015\}$ in rows from top to bottom. When scientific notation is used for the y-axis values in certain subplots, it is denoted by a number above the respective y-axis.}
  \label{fig:adjust-reallocation-cost-utility}
\end{figure}

In Figure~\ref{fig:adjust-reallocation-cost-utility} we modulate the cost of reallocation, $\eta$. We see that higher $\eta$ values lead to higher optimal $\tau$ values for both OIRA and ODRA strategies. This is to be expected, for although low $\tau$ values might lead to higher gains in fees, this also leads to more frequent resets which in turn come with a higher cost. 

\section{Conclusion}
\label{sec:conclusion-new}
This paper fills existing gaps in the literature regarding strategic liquidity provision strategies for LPs in Uniswap v3. Whereas earlier  work has either optimized for complex liquidity positions in static environments, or simple positions with dynamic reallocations, our work simultaneously provides complex, context-dependent liquidity allocations that dynamically reallocate as prices evolve in v3 contracts. Our results show that such liquidity provision strategies provide large gains for LPs relative to baseline uniform allocation strategies (similar to liquidity allocation in Uniswap v2) in multiple economic environments for decentralized exchanges. Natural directions for future work include: incorporating LVR~\cite{milionis2022automated} into the objective definition to optimize the strategy of a delta-hedge LP, incorporating a game-theoretic framework to liquidity provision which is more apt for large LPs, and modelling competition between different pools such as v2 and v3 pools for the same token pairs. 



\bibliography{refs.bib}

\begin{thebibliography}{10}

\bibitem{uniswapv1white}
Hayden Adams.
\newblock Uniswap whitepaper, 2018.
\newblock URL: \url{https://hackmd.io/@HaydenAdams/HJ9jLsfTz}.

\bibitem{adams2020uniswap}
Hayden Adams, Noah Zinsmeister, and Dan Robinson.
\newblock Uniswap v2 core, 2020.
\newblock URL: \url{https://uniswap.org/whitepaper. pdf}.

\bibitem{adams2021uniswap}
Hayden Adams, Noah Zinsmeister, Moody Salem, River Keefer, and Dan Robinson.
\newblock Uniswap v3 core, 2021.
\newblock URL: \url{https://uniswap.org/whitepaper-v3.pdf}.

\bibitem{angeris2020improved}
Guillermo Angeris and Tarun Chitra.
\newblock Improved price oracles: Constant function market makers.
\newblock In {\em Proceedings of the 2nd ACM Conference on Advances in
  Financial Technologies}, pages 80--91, 2020.

\bibitem{angeris2020does}
Guillermo Angeris, Alex Evans, and Tarun Chitra.
\newblock When does the tail wag the dog? {Curvature and market making}.
\newblock {\em arXiv preprint arXiv:2012.08040}, 2020.

\bibitem{angeris2019analysis}
Guillermo Angeris, Hsien-Tang Kao, Rei Chiang, Charlie Noyes, and Tarun Chitra.
\newblock An analysis of {Uniswap} markets.
\newblock {\em arXiv preprint arXiv:1911.03380}, 2019.

\bibitem{aoyagi2020lazy}
Jun Aoyagi.
\newblock Lazy liquidity in automated market making.
\newblock {\em Available at SSRN 3674178}, 2020.

\bibitem{arrow1965aspects}
Kenneth~Joseph Arrow.
\newblock {\em Aspects of the theory of risk-bearing}.
\newblock Helsinki, 1965.

\bibitem{bar2023uniswap}
Yogev Bar-On and Yishay Mansour.
\newblock Uniswap liquidity provision: An online learning approach.
\newblock {\em arXiv preprint arXiv:2302.00610}, 2023.

\bibitem{capponi2021adoption}
Agostino Capponi and Ruizhe Jia.
\newblock The adoption of blockchain-based decentralized exchanges.
\newblock {\em arXiv preprint arXiv:2103.08842}, 2021.

\bibitem{cartea2022decentralised}
{\'A}lvaro Cartea, Fay{\c{c}}al Drissi, and Marcello Monga.
\newblock Decentralised finance and automated market making: Predictable loss
  and optimal liquidity provision.
\newblock {\em Available at SSRN 4273989}, 2022.

\bibitem{evans2020liquidity}
Alex Evans.
\newblock Liquidity provider returns in geometric mean markets.
\newblock {\em arXiv preprint arXiv:2006.08806}, 2020.

\bibitem{evans2021optimal}
Alex Evans, Guillermo Angeris, and Tarun Chitra.
\newblock Optimal fees for geometric mean market makers.
\newblock {\em arXiv preprint arXiv:2104.00446}, 2021.

\bibitem{fan2022differential}
Zhou Fan, Francisco J.~Marmolejo Coss{\'{\i}}o, Ben Altschuler, He~Sun, Xintong
  Wang, and David~C. Parkes.
\newblock Differential liquidity provision in {Uniswap v3} and implications for
  contract design.
\newblock In {\em 3rd {ACM} International Conference on {AI} in Finance,
  {ICAIF}}, pages 9--17, 2022.

\bibitem{frongillo2023axiomatic}
Rafael Frongillo, Maneesha Papireddygari, and Bo~Waggoner.
\newblock An axiomatic characterization of {CFMMs} and equivalence to
  prediction markets.
\newblock {\em arXiv preprint arXiv:2302.00196}, 2023.

\bibitem{goyal2022finding}
Mohak Goyal, Geoffrey Ramseyer, Ashish Goel, and David Mazi{\`e}res.
\newblock Finding the right curve: Optimal design of constant function market
  makers.
\newblock {\em arXiv preprint arXiv:2212.03340}, 2022.

\bibitem{heimbach2022risks}
Lioba Heimbach, Eric Schertenleib, and Roger Wattenhofer.
\newblock Risks and returns of {Uniswap v3} liquidity providers.
\newblock {\em arXiv preprint arXiv:2205.08904}, 2022.

\bibitem{charmalphavault}
Max.
\newblock Introducing {Alpha Vaults}--an {LP} strategy for {Uniswap V3}, 2021.
\newblock URL:
  \url{https://medium.com/charmfinance/introducing-alpha-vaults-an-lp-strategy-for-uniswap-v3-ebf500b67796}.

\bibitem{milionis2023myersonian}
Jason Milionis, Ciamac~C Moallemi, and Tim Roughgarden.
\newblock A {Myersonian} framework for optimal liquidity provision in automated
  market makers.
\newblock {\em arXiv preprint arXiv:2303.00208}, 2023.

\bibitem{milionis2022automated}
Jason Milionis, Ciamac~C Moallemi, Tim Roughgarden, and Anthony~Lee Zhang.
\newblock Automated market making and loss-versus-rebalancing.
\newblock {\em arXiv preprint arXiv:2208.06046}, 2022.

\bibitem{neuder2021strategic}
Michael Neuder, Rithvik Rao, Daniel~J Moroz, and David~C Parkes.
\newblock Strategic liquidity provision in uniswap v3.
\newblock {\em arXiv preprint arXiv:2106.12033}, 2021.

\bibitem{pratt1978risk}
John~W Pratt.
\newblock Risk aversion in the small and in the large.
\newblock {\em Econometrica}, 32:122--136, 1964.

\bibitem{schlegel2022axioms}
Jan~Christoph Schlegel, Mateusz Kwa{\'s}nicki, and Akaki Mamageishvili.
\newblock Axioms for constant function market makers.
\newblock {\em Available at SSRN}, 2022.

\bibitem{tassy2020growth}
Martin Tassy and David White.
\newblock Growth rate of a liquidity provider’s wealth in $xy= c$ automated
  market makers, 2020.

\bibitem{uniswapsfinancialalchemy}
Dave White, Martin Tassy, Charlie Noyes, and Dan Robinson.
\newblock Uniswap's financial alchemy, 2020.
\newblock URL: \url{https://research.paradigm.xyz/uniswaps-alchemy}.

\bibitem{zhao2021understand}
Wenqi Zhao, Hui Li, and Yuming Yuan.
\newblock Understand volatility of {Algorithmic Stablecoin: Modeling,}
  verification and empirical analysis.
\newblock In {\em Financial Cryptography and Data Security}, volume 12676 of
  {\em Lecture Notes in Computer Science}, pages 97--108. Springer, 2021.

\end{thebibliography}

\appendix

\section{Casting OIRA Variants as a Convex Optimization Problem}
\label{appendix:convex-optimization}

In this section we consider a slightly different reallocation cost model and risk-aversion utility function for LPs. In this framework we are able to cast the problem of computing a natural variant of the optimal context-independent $\tau$-reset allocation (which we denote OIRA') as a convex optimization problem. 

\paragraph*{Modified Reallocation Cost}
In Section \ref{sec:dynamic-liquidity-provision} we modeled reallocation costs as being proportional to the capital an LP locks in the contract for a given epoch. For this section, we make the simplifying assumption that reallocation costs are simply proportional to the overall wealth accumulated by the LP at the time of a liquidity reset. In more detail, let us suppose that the LP triggers a reset at time $t$, using their overall earnings, $W_t$, to mint a proportional allocation $\vect{x}$ given the contract-market price $P_t$. Reallocation costs are once more parametrized by a single parameter $\eta \in [0,1]$, and we assume that the LP pays $(1-\eta)W_t$ in reallocation costs and applies the proportional allocation $\vect{x}$ to be used with available funds $\eta W_t$. 

As before, the price sequence $\vect{P}$ is partitioned into epochs when a given reset strategy is applied. Let us consider the $j$-th epoch, denoted by $E^j$. The epoch has $W^j$ total earnings accrued at its beginning, and the LP uses $\vect{x}^j$ as a proportional allocation for the duration of the epoch. Given the cost model from the previous paragraph, it follows that the LP's wealth at the end of the epoch (and hence at the beginning of $E^{j+1}$) is given by: 
$$
W^{j+1} = \eta W^j \cdot V(\vect{x}^j,E^j).
$$

Now let us suppose that applying a dynamic reset strategy $\Lambda$ to $\vect{P}$ results in $k$ epochs. We can express the overall earnings of the LP as follows: 
$$
V(\Lambda,W,\vect{P}) = \eta W^k \cdot V(\vect{x}^j,E^j) = W \left( \eta^k \prod_{i=1}^k V(\vect{x}^j,E^j) \right),
$$
where we have unpacked each $W^j$ in the product. Importantly, we notice that this expression is in fact linear in $W$, hence it follows that $V(\Lambda,W,\vect{P}) = W \cdot V(\Lambda,\vect{P})$. 

\paragraph*{Logarithmic Risk-aversion}
In Section \ref{sec:optimal-tau-reset-strategies} we modeled risk-averse LPs via Constant Absolute Risk Aversion (CARA) utilities. Another common choice for risk aversion in LPs is that of logarithmic utility, as in the work of \cite{cartea2022decentralised}.  

\begin{definition}[Logarithmic Utility]
The logarithmic utility function $u_l:\mathbb{R} \rightarrow \mathbb{R}$ is:
$$
u_l(x) = \log (x)
$$
\end{definition}

\paragraph*{Optimal Context-Independent Allocations as a Convex Optimization Problem}

In this section we focus on context-independent $\tau$-reset allocations. We recall that such dynamic liquidity provision strategies have the constraint that the LP uses the same relative allocation at each reset (i.e. all $\vect{x}^j$ are equal in all epochs), hence the space of all such allocations can be identified with the convex $(2\tau + 2)$-dimensional simplex, with $\vect{x} \in \Delta^{2\tau + 2}$ denoting the common strategy used at all liquidity resets. We let $\Lambda(\vect{x})$ denote the context-independent $\tau$-reset allocation that makes use of $\vect{x}$ at each reset.  

Following the notation of Section \ref{sec:optimal-tau-reset-strategies}, we are ultimately interested in optimizing the following objective:
$$
V^{u_l}_{\tau,\cP}(\Lambda(\vect{x})) = \mathbb{E}_{\vect{P} \sim \cP} [u_l(V,\Lambda(\vect{x}),\vect{P})]
$$
where $\cP$ is a liquidity-independent price distribution and $u_l$ is the logarithmic risk-aversion utility function.

\begin{theorem}\label{thm:concave-OIRA}
    $V^{u_l}_{\tau,\cP}$ is concave in $\vect{x}$.
\end{theorem}
\begin{proof}
    From the previous section, we know that $V(\Lambda(\vect{x}),\vect{P}) = \eta^k \prod_{i=1}^k V(\vect{x},E^j)$. If we apply logarithmic utilities to this expression, we obtain:
    $$
    u_l(V(\Lambda(\vect{x}),\vect{P})) = k\log(\eta) + \sum_{i=1}^k \log(V(\vect{x},E^j)).
    $$
    We recall that for each epoch, $V(\vect{x},E^j)$ is linear in $\vect{x}$. Since the $\log$ function is concave and nondecreasing, it follows that $\log(V(\vect{x},E^j))$ is in turn concave in $\vect{x}$. It follows that $u_l(V(\Lambda(\vect{x}),\vect{P}))$ is concave in $\vect{x}$ since it is the sum of a constant independent of $\vect{x}$ and another sum of concave functions in $\vect{x}$. 
\end{proof}

As before, we can approximate the objective function by taking samples from $\cP$. As such, suppose that $\vect{P}_1,\dots,\vect{P}_N \sim \cP$. The empirical average earnings of an LP are given by: 
$$
\hat{V}^{u_l}_{\tau,\cP}(\vect{x} \mid \vect{P}_1,\dots,\vect{P}_N) = \frac{1}{N} \sum_{q=1}^N u_l(V(\Lambda(\vect{x}),\vect{P}_q)), 
$$
where this expression is concave in $\vect{x}$ as per Theorem \ref{thm:concave-OIRA}. If we take expectations we obtain 
$$
\mathbb{E}_{\vect{P}_1,\dots,\vect{P}_N \sim \cP} \left[ \hat{V}^{u_l}_{\tau,\cP}(\vect{x} \mid \vect{P}_1,\dots,\vect{P}_N) \right] = V^{u_l}_{\tau,\cP}(\Lambda(\vect{x}).
$$
Ultimately, we denote the optimal liquidity allocation in this context by OIRA', and the convex optimization problem used to compute it is as follows: 
\begin{equation}
\begin{aligned}
\min_{\vect{x}} \quad & -\hat{V}^{u_l}_{\tau,\cP}(\vect{x} \mid \vect{P}_1,\dots,\vect{P}_N)\\
\textrm{s.t.} \quad & \vect{x} \in \Delta^{2\tau + 2}\\
\end{aligned}
\end{equation}

\newpage
\begin{table} [bp]
  \label{tab:notation}

  \begin{tabular}{ll}
    \hline
    \textbf{Symbol} & \textbf{Description} \\
    \hline
    $\gamma$  & Fee tier of contract\\
    $\vect{\mu} = \{B_{-n},\dots,B_{m}\}$  & Set of price buckets in contract\\
    $B_i = [a_i,b_i]$  & $i$-th price bucket\\
    $P \in (0,\infty)$  & Contract price\\
    $\cV^{(3)}(L,P,B_i)$ & Token bundle value of $L$ units of $B_i$-liquidity at contract price $P$\\
    $(\sigma,\vect{L})$ & Liquidity Allocation profile of v3 contract\\
    $\vect{P} = (P_1,\dots,P_T)$ & Contract-market price sequence over time horizon of length $T$\\
    $P_t = (P_{c,t},P_{m,t})$ & Contract-market price at time $t$\\
    $W$ & Initial token $B$ budget of LP\\
    $\vect{x} = (x_{-n},\dots,x_m)$ & Proportional liquidity allocation\\
    $\cB((z_1,z_2),P_m)$ & Token $B$ value of $(z_1,z_2)$ bundle of $A$ and $B$ tokens at market price $P_m$\\
    $\vect{\ell} = (\ell_{-m},\dots,\ell_n)$ & Absolute liquidity allocation\\
    $w_i, w_i'$ & Token $B$ value of 1 unit of liquidity at beginning (end resp.) of time horizon\\
    $F^A(\vect{x},W,\vect{P})$, $F^B(\vect{x},W,\vect{P})$  & Token $A$ and $B$ fee rewards for $\vect{x}$ over $\vect{P}$ with initial budget $W$\\
    $C(\vect{x},W,\vect{P})$ & Token $B$ value of LP position at time $T$ \\
    $V(\vect{x},W,\vect{P})$  & Token $B$ earnings of LP at time $T$\\
    $\eta$ & Reallocation cost\\
    $(E^1,\dots,E^k)$ & Epochs of reset LP strategy\\
    $\vect{t} = (t^1,\dots,t^k)$ & Reset times over contract-market price sequence\\
    $W^j$ & Cumulative token $B$ wealth at beginning of $E^j$\\
    $C^j$ & Context feature used as input of NN at beginning of $E^j$\\
    $\matr{X}$ & Matrix with proportional allocations at each epoch (rows)\\
    $(\vect{t},\matr{X})$ & Realized dynamic liquidity provision allocation profile\\
    $\Lambda$ & Reset LP strategy\\
    $V(\Lambda,\vect{P})$ & Token $B$ earnings of LP with $W = 1$ under reset LP strategy $\Lambda$\\
    $\cP$ & Distribution over contract-market prices sequences\\
    $R$ & Number of rounds of market-price updates\\
    $\vect{k} = (k_1,\dots,k_R)$ & Vector with number of non-arbitrage trades per round\\
    $\vect{\lambda} = (\lambda_1,\dots,\lambda_R)$ & Vector with magnitude of non-arbitrage trades per round\\
    $\bar{\lambda}, \alpha$ & Average and spread of values in $\vect{\lambda}$ over all rounds\\
    \hline
  \end{tabular}
  \caption{Relevant Notation}
\end{table}

\end{document}